\documentclass[10pt, conference, letterpaper]{IEEEtran}

\usepackage[dvipsnames]{xcolor}
\usepackage{cite}
\usepackage[T1]{fontenc} 
\usepackage{amssymb,amsmath,amsthm,amsfonts}
\usepackage{mathabx}
\usepackage[normalem]{ulem} 
\usepackage{cleveref}

\crefformat{section}{\S#2#1#3} 
\crefformat{subsection}{\S#2#1#3}
\crefformat{subsubsection}{\S#2#1#3}
\usepackage{url}
\usepackage{bm}
\usepackage{caption}
\usepackage[linesnumbered,ruled,vlined,noend]{algorithm2e}
\makeatletter
\newcommand{\nosemic}{\renewcommand{\@endalgocfline}{\relax}}
\newcommand{\dosemic}{\renewcommand{\@endalgocfline}{\algocf@endline}}
\let\oldnl\nl
\newcommand{\nonl}{\renewcommand{\nl}{\let\nl\oldnl}}
\makeatother

\SetCommentSty{mycommfont}
\SetKwComment{Comment}{$\triangleright$\ }{\hspace{-0.7cm}}
\usepackage{graphicx}
\graphicspath{{img/}}
\usepackage{color,soul}
\usepackage{booktabs} 
\usepackage{multirow} %
\usepackage{accents}
\usepackage{textcomp}
\usepackage{diagbox}

\usepackage[font=small]{caption}
\usepackage[labelformat=simple,font=footnotesize]{subcaption}

\usepackage{lastpage}
\usepackage{setspace}
\usepackage[inline]{enumitem}
\newcommand{\eg}{\textit{e.g.},}
\newcommand{\ie}{\textit{i.e.},}
\newcommand{\etc}{\textit{etc}.\@}

\newcommand{\cat}[1]{\smallskip\noindent\textbf{#1.}}

\newtheoremstyle{example}{}{}{}{}{\bfseries}{\smallskip.}{ }{}
\theoremstyle{example}

\SetAlgorithmName{Alg.}{alg.}{List of Algorithms}

\newcommand{\pslb}{\textsmaller{\textsf{PRIME}}}

\newcommand{\ecmp}{\text{ECMP}}
\newcommand{\ecn}{\text{ECN}}
\newcommand{\fct}{\text{FCT}}
\newcommand{\reps}{\textsmaller{\textsf{REPS}}}
\newcommand{\strack}{\textsmaller{\textsf{STrack}}}
\newcommand{\rr}{\text{RR}}
\newcommand{\ar}{\textsmaller{\textsf{AR}}}
\newcommand{\ev}{\text{EV}}
\newcommand{\mev}{\text{MP-EV}}
\newcommand{\ack}{\text{ACK}}
\newcommand{\nack}{\text{NACK}}
\newcommand{\lb}{\text{LB}}
\newcommand{\spa}{\text{SP}}
\newcommand{\obps}{\text{Congestion Oblivious-PRIME}}
\newcommand{\ob}{\textsmaller{\textsf{CO-PRIME}}}

\begin{document}
\bstctlcite{IEEEexample:BSTcontrol}

\title{PRIME: Pseudo-Random Integrated Multi-Part Entropy for Adaptive Packet Spraying in AI/ML Data centers}

\author{Ashkan Sobhani, Sogand Sadrhaghighi, Xingjun Chu}

\maketitle

\begin{abstract}
Large-scale distributed training in production data centers place significant demands on network infrastructure. In particular, significant load balancing challenges arise when processing AI/ML workloads, consisting of low-entropy, bursty and long-lived flows. Existing solutions designed for Ethernet, such as Equal-Cost Multi-Path (\ecmp) struggle to maintain high network utilization. While major industry players (\eg\ Ultra Ethernet Consortium) and parts of academia have proposed \textit{packet spraying} to enhance AI/ML workload performance, we argue that existing packet spraying solutions lead to buffer inflation over time, negatively affecting network performance. Specifically, when \ack\ coalescing is used, these solutions lead to stale information, degrading network performance. Additionally, in asymmetric network conditions- such as mix of ordered an unordered traffic, or link degradation and failures- existing packet spraying solutions often lead to increased tail latency. In this paper, we present the design and evaluation of \pslb, a pseudo-randomized round-robin approach to packet spraying that considers the network topology to optimize load distribution and performance. \pslb\ uses congestion as an indicator to re-balance the load. To this extent, \pslb\ takes into account various congestion signals, accounting for congestion severity, and their decay times to avoid network hotspots. We extensively evaluated \pslb\ using large-scale production-level simulator. Our results indicate that, compared to existing solutions, \pslb\ leads to up to $15\%$ improvement for permutation traffic and up to $27\%$ improvement in network degradation scenarios.
\end{abstract}

\section{Introduction}
\cat{Motivation}
The rapid growth in AI's computational demands necessitates distributed training to efficiently scale across large GPU clusters and leverage parallel processing for faster training time~\cite{addanki2024challenging}. To speed up the training process significant research has been conducted in various areas, \eg\ parallelization strategies~\cite{lu2017flexflow}, topology optimization~\cite{hoefler2022hammingmesh}, \etc\ Communication, however, is the main bottleneck in distributed training~\cite{zhang2020network}, with  congestion 
and unbalanced load being the main culprit behind slow downs. 

While congestion control and load-balancing are two of the most widely studied problems in the context of data centers~\cite{alizadeh2014conga,alizadeh2010data,smartttreps2024}, AI/ML workloads have different characteristics compared to traditional data center workloads~\cite{hoefler2023datacenterethernetrdmaissues}, requiring different congestion control and load-balancing mechanisms. In particular, traditional data center traffic features a large number of short-lived, small data flows (known as mice flows) with diverse destinations and random start/stop time~\cite{alizadeh2010data}. On the other hand, AI/ML traffic consists of fewer long-lived, large data flows (known as elephant flows) that start synchronously and follow specific patterns, such as all-reduce or all-to-all communication~\cite{aft-ai-traffic-00,ouyang2021communication}. As such, while traditional data traffic exhibits high entropy due to randomness, AI/ML workloads has lower entropy with predictable flow patterns. Consequently, optimizing job completion time in AI/ML training requires minimizing tail flow completion time (\fct) by using an effective load balancing mechanism to prevent resource bottlenecks and optimize GPU utilization in data centers~\cite{cao2024network}. 


%

Equal-Cost Multi-Path (\ecmp) routing~\cite{thaler2000multipath} is a standard flow-level load balancing algorithm, which randomly assigns flows to different paths. While \ecmp\ is widely adopted in traditional data centers, it suffers from hash collisions and is not able to adapt to network asymmetry (\eg\ link failure), both of which increase the tail latency of AI/ML workloads. 
To address these limitations, some works, \eg\ CONGA~\cite{alizadeh2014conga} and Letflow~\cite{vanini2017let}, split a flow into smaller bursts, called flowlets, and dynamically route them across multiple paths. These mechanisms, however, are not flexible as the flowlet gap is set to a fixed value~\cite{huang2021mitigating}, affecting the path utilization.  Additionally, regardless of the design merits, it is expensive and time-consuming to adopt schemes that require new switch processing, and impractical to rely on designs with centralized controllers (\eg\ Hedera~\cite{259355}). 

In the wake of this problem, both academia and major industry players, \eg\ Ultra Ethernet Consortium~\cite{ultraethernet,metz2024empowering}, adopt packet spraying to enhance AI/ML workload performance.
Packet spraying reduces collisions by distributing packets of every flow across multiple paths, thereby lowering the likelihood of congestion on any single path and enhancing overall network efficiency for AI/ML traffic~\cite{dixit2013impact}. Nonetheless, random packet spraying results in bursts and queue build up, increasing tail latency. To mitigate these bursts, some approaches, employ Adaptive Routing (\ar) in which packets are sprayed based on switches' local queue information\cite{ghorbani2017drill}. \ar\ solutions, however, suffer from collisions on higher tiers of the network, since switches decide on the path locally without having visibility over the entire path. To account for the entire path, some studies, such as \strack~\cite{le2024strack} and \reps~\cite{smartttreps2024}, distribute packets across different paths based on entropy values (\ev s) derived from a hash function applied to the flow identifier. \strack, however, suffers from stale congestion information and over penalizing the congested paths. The former is due to \strack\ not using any decay functions to account for drainage of buffers over time and the latter is due to using accumulative penalties. \reps\ addresses these limitations by employing time-based decay function and non-cumulative penalties. \reps, however, cannot generate perfectly uniform \ev s, leading to buffer inflation over time. Additionally, \reps\ effectiveness is limited when \ack\ coalescing is used, as fewer \ack s reduce the granularity of congestion signals, leading to stale congestion information and load imbalance in the network. Finally, \reps\ suffers from performance degradation in the presence of network asymmetry. 

\cat{Our approach}
AI/ML workloads requires a \lb\ solution that sustain high network utilization under various scenarios and is robust to \ack\ coalescing. 
In particular, the problem we are addressing in this work is, \textbf{\textit{how to have a robust load balancing mechanism that maintains high network utilization for AI/ML workloads while remaining resilient to reduced congestion visibility caused by \ack\ coalescing}} by proposing \pslb, a load balancing approach that sprays packets of a flow across different available paths. \pslb\ is a hybrid switch-host approach, however, unlike complex in-network load balancing implementations, it does not require any switch support besides similar support to \ecmp. On the host-side, \pslb\ is simple and can be implemented in the NIC hardware with minimal memory/area footprint. \pslb's design is based on three main insights. Firstly, \ev s can be generated based on the network topology in a partially deterministic and controlled randomized manner. Secondly, when congestion is detected on a certain path, the path shouldn't be taken by any packets until the congestion is resolved. Thirdly, an effective load balancing solution considers not only the severity of congestion but also the time congestion signals were received, avoiding stale information in path selection.

To increase the scalability of the proposed \lb, \pslb\ generates a multi-part \ev\ (\mev), where each part is an \ev\ that specifies the index of the next-hop switch uplink port. The number of parts depends on the network topology, \eg\ a $3$-tier FatTree uses a \mev\ with two parts, while a $2$-tier FatTree topology uses one-part \mev. Using a \mev, \pslb\ distributes the traffic evenly across available links in each tier of the network and reduces the space required for \ev s by precisely accounting for the exact number of paths. For each \ev\ in \mev, \pslb\ employs a pseudo-randomized round-robin algorithm to iterate through uplink ports and distribute traffic evenly in the network. To add randomness and avoid congestion on a single path, \pslb\ shuffles each part of \mev\ after an iteration is finished, \ie\ considers different permutation of an \ev. We argue that \pslb's approach for generating \ev s not only reduces complexity on the hosts by eliminating the need for hashing but also distributes traffic more uniformly across available paths, resulting in less collisions. One important feature of \pslb\ is that it uses congestion signals to dynamically steer traffic away from paths that are experiencing hotspots. To this extent, \pslb\ interprets various signals differently, accounting for the severity of congestion. For example, receiving a \nack\ packet indicates a more severe congestion status compared to receiving an \ecn-marked packet, and as such \pslb\ penalizes those paths differently. Additionally, \pslb\ uses a decay function to account for the timing of congestion events, prioritizing recent signals and preventing the use of stale information. We argue that \pslb's approach in using congestion history mitigates the challenge of using stale information, specifically when \ack\ coalescing is used. Our results further emphasizes that \pslb\ significantly reduces the flow completion time (\fct) across various traffic patterns in multiple scenarios, including asymmetric network conditions.

\cat{Contributions} Our contributions in this paper are:
\begin{itemize}
	\item We present the design and evaluation of \pslb, a load balancing scheme that sprays packets of a flow across different available paths. \pslb\ is a hybrid host-switch design that requires minimal modifications to existing switches, making it compatible with currently available switch hardware. Additionally, \pslb\ simplifies the operation and space needed on the host stack. 
	\item To accommodate the growing scale of AI/ML data centers, we developed multi-part \ev\ (\mev), where the number of parts are decided based on the topology tier number. 
	\item Our host algorithm employs pseudo-randomized round-robin to enforce uniform distribution of packets and avoid collisions in the network core. Our proposed algorithm considers different congestion signals and their timing to avoid paths experiencing hotspots.
	\item As a proof of concept, we have implemented  \pslb\ in a production-level proprietary packet-level simulator and evaluated it in various scenarios and under different traffic patterns. Our results indicate that \pslb\ outperforms alternative packet spraying approaches in all scenarios and under all traffic patterns~\cite{smartttreps2024}.
\end{itemize}
\section{Background}
This section covers background used in \pslb\ design. Specifically, we first cover the main congestion signals used in the design and evaluation of \pslb. Following that we introduce key terms and concepts regarding to load balancing.

\cat{Congestion control algorithms} Congestion control algorithms rely on congestion signals to adjust the flows' sending rate or window, with the goal of maintaining high network utilization. The following are the main congestion signals used in AI data center networks: 
\begin{itemize}
	\item \textit{ECN-marking}: With \textit{Explicit Congestion Notification} (\ecn), switches set a bit in the traffic class field of the IP header when a packet experiences congestion. Different policies can be employed to decide if a packet must be marked. For example, in Random Early Detection (RED)~\cite{floyd1993random}, switches randomly mark packets with a probability linear in the switch queue size, if the queue size is within two thresholds $(K_{min},K_{max})$. 
	
	\item \textit{Packet loss and trimmed packets}: Packet loss is a key indicator of severe congestion in the network. However, to meet the strict latency of AI/ML workloads, packet trimming is used to avoid silent packet drops by quickly detecting and reacting to congestion~\cite{handley2017re}. In particular, in packet trimming, a switch removes parts of the packet (\eg\ packet payload) rather than dropping it. When a receiver node, receives a trimmed packet, it sends a negative acknowledgment (\nack) to the sender indicating which packet was trimmed. Since packet trimming keeps essential information of the packet (\eg\ packet header) intact, the sender can quickly detect and react to congestion.
\end{itemize} 

\cat{Entropy value} Entropy value (\ev) is a value set in the packet header that determines the path, the packet takes, through a network. In particular, \ev\ is set by the sender, and is used as an input to a hash function in switches specifying the switch output port. Possible header fields that can be used to carry the \ev\ are the Source Port field in the packet header~\cite{lu2018multi} or the Flow Label in IPv6~\cite{qureshi2022plb}. \pslb\ leverages \ev s for efficient load balancing without requiring the exact mapping between packet's \ev\ and its resulting path.

\cat{Packet spraying} It is well-known that \ecmp\ routing causes congestion in the core of the network due to hash collisions~\cite{9269053}. On the other hand, random packet spraying, randomly distributes individual packets across all available paths between a sender and a receiver~\cite{dixit2013impact}. This process is done by selecting a random \ev\ for every packet at a sending host. Random packet spraying, however, may result in arbitrary-length queue build up and large latency in the network. This behavior of packet spraying can be theoretically proved using balls-into-bins model~\cite{reps2025}.
\section{Design}
\subsection{Design Motivation} 
\begin{figure}[!t]
	\centering
	\includegraphics[width=\columnwidth]{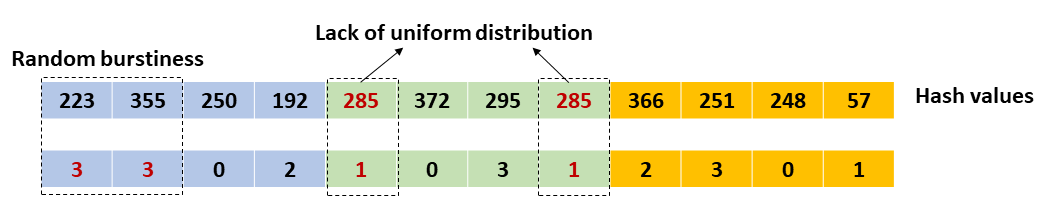}
	\caption{Example of hash function limitation in \ev\ generation.}
	\label{hash_limitation}
\end{figure}

The design of \pslb\ is motivated by the following observations in production data center networks:

\cat{Non-uniform load in the core}  It is crucial to avoid collisions in the network core. To this extent, when using packet spraying, it is essential that in a switch with $L$ uplink ports, no two packets are assigned to the same port, \ie\ the same \ev. This is affected by \ev\ generation, which can be done using various approaches, such as using hash functions~\cite{dixit2013impact}. Hash functions, while simple to implement, do not provide uniform usage of upstream ports~\cite{cao2013per}. The reason behind this non-uniform usage is that hash functions cannot produce perfectly uniform values, resulting in some ports being used more frequently than others. To address this limitation, the output value space of the hash functions can be expanded, albeit with increased computational expense. Nonetheless, regardless of the hash output size, its value needs to be mapped to a smaller space corresponding to port numbers. Such mapping, however, can lead to random bursts when a port is used more often than other ports in a short time. Fig.~\ref{hash_limitation} depicts an illustration of the aforementioned problems. 

\cat{Flow completion time} Most AI training frameworks rely on collective communication operations, where the overall collective time is determined by the slowest participant. As such, the primary concern in distributed training workloads is the collective completion time, specified by the tail latency. This means that even if most flows are finished quickly, the overall performance is bottle necked by the stragglers. 

\begin{figure}[!t]
	\centering
	\includegraphics[width=0.95\columnwidth]{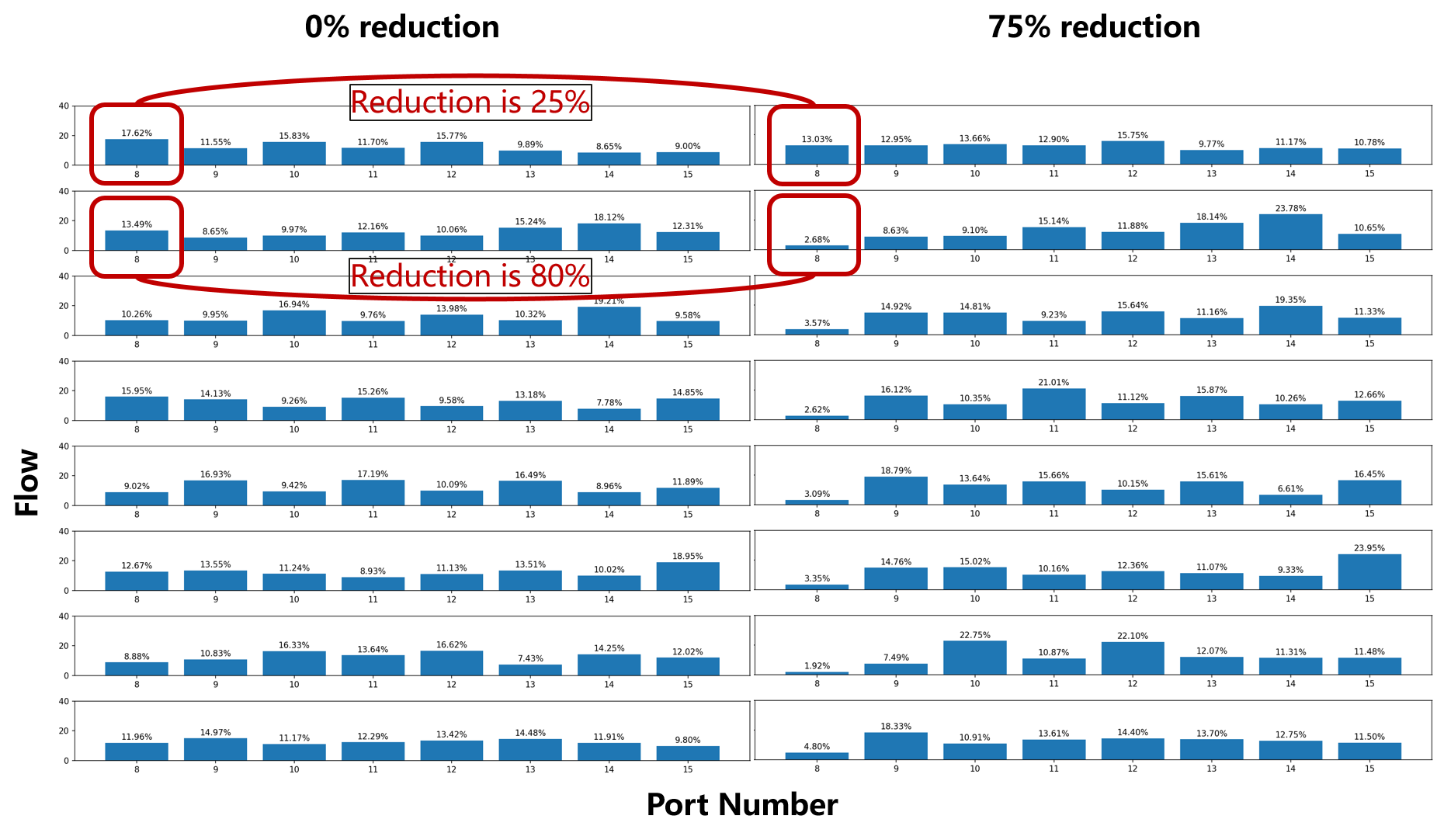}
	\caption{Load distribution of $8$ flows across different ports of a leaf switch under varying degradation levels when using \reps.}
	\label{assymetry}
\end{figure}

\cat{Unfair penalization of flows during asymmetric-network conditions} Unlike traditional networks, where fairness ensures equal resource distribution, distributed training relies on synchronized gradient updates. As such, the overall performance of distributed training is determined by the tail latency rather than fairness. Although scheduling flows in any order while maintaining uniform load balancing across the network core may be sufficient to reduce tail latency, certain scheduling orders can disproportionately penalize some flows when asymmetric network conditions, \eg\ link degradation, occurs.  
To show this effect we have implemented REPS~\cite{smartttreps2024} in a $2$-tier Fattree topology with $128$ servers, $15$ leaf switches and $7$ core switches. The bandwidth of all links are set to $400$~Gbps. $18$ flows traverse from servers connected to leaf switch~$0$ to leaf switch~$1$. We then introduce one link to be degraded from $0$\% to $75$\% and measure the aggregated load on the uplinks of leaf switch~$0$. Fig.~\ref{assymetry} shows the aggregate packets distributed across different ports that traverse leaf switch~$0$. We can observe that when no link is degraded the aggregated load is distributed evenly across all ports. However, by introducing degradation, the aggregated load sent to the degraded link reduces proportionally to the degradation ratio. However, the load distribution across different flows is uneven. In particular, packets from degraded link are redirected to non-degraded links. However, the proportion of shifted packets varies across different flows. For example, with $75\%$ degradation, the load of flow~$2$ on port~$1$ is much less than the load of flow~$1$. As such, certain ports handle disproportionate share of packets, highlighting the load imbalance.

\cat{Inaccurate network feedback integrity in presence of \ack\ coalescing}
Dynamically adjusting traffic distribution based on real-time congestion signals ensures efficient resource utilization and steers traffic away from congested paths in the network. Load balancing schemes can leverage both positive (\ie\ reinforcing traffic distribution decision) and negative (\ie\ indicating congestion) feedback, but their effectiveness hinges on the timeliness of that feedback. This is specifically important when \ack\ coalescing is used, which can delay the signal and lead to outdated decisions. 
\subsection{Design Requirements}
\begin{figure}[!t]
	\centering
	\includegraphics[width=\columnwidth]{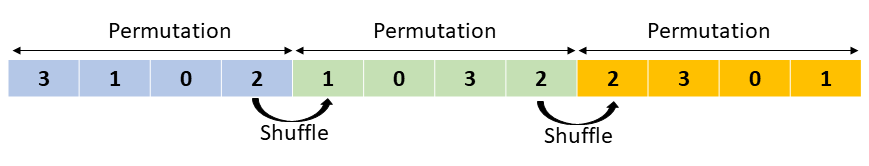}
	\caption{Example of using pseudo-randomized round-robin for generating \ev s.}
	\label{round_robin}
\end{figure}

We have designed \pslb\ such that the following requirements are satisfied:

\cat{Uniform usage of all available upstream ports} \pslb\ balances the load in the network core by ensuring that all available upstream ports are used uniformly. \pslb\ achieves this by employing a \textit{pseudo-randomized round robin} \ev\ generation algorithm, where each host uses a different order for port selection, reducing synchronization between packets of different flows. Additionally, after each full cycle of round-robin assignments, the order of the port numbers is randomly reshuffled to prevent any potential synchronization that could occur between packets from different flows (see Fig.~\ref{round_robin}).

\cat{Congestion-aware path selection using congestion history} \pslb\ uses congestion history to avoid congested paths. Different congestion signals indicate varying levels of congestion severity. For example, \textit{explicit congestion notification} (\ecn) specifies an early congestion, while a packet drop shows a more severe congestion. As such, \pslb\ distinguishes between different congestion signals and interprets them based on their severity. In particular, \pslb\ looks into the following congestion signals:
\begin{itemize}
	\item \textit{ECN-marking}: 
	While \ecn\ was originally designed to mark packets at the time of enqueue~\cite{ramakrishnan2001addition}, it has been shown that \ecn\ marking on packets at the time of dequeue allows congestion control algorithms to react faster. Additionally, dequeue \ecn\ marking can be easily implemented on most switches~\cite{wu2012tuning}. As such, \pslb\ assumes switches use RED and mark \ecn\ packets at dequeue. \pslb\ penalizes the path associated with an \ecn-marked packet for a short period of time, allowing for transient congestion to pass.
	
	
	\item \textit{Packet loss and trimmed packets}: Trimmed packets and packet loss initiate a \nack\ packet to be sent from the receiver to the sender, indicating that the network is experiencing severe congestion. To account for this severeness, \pslb\ penalizes the path associated with a \nack\ packet for an extended period of time.          
\end{itemize} 

\subsection{\pslb\ Operation}
 In the following, \pslb's procedure is described in more details:

\cat{Process at hosts} For the first Bandwidth-Delay Product (\textit{bdp}) of packets, there is no congestion history available. As such, \pslb\ explores a list of all possible \ev s and generates a \mev\ based on the network topology (see Subsection~\ref{design_elements}). For the remainder of the flow's lifetime, \pslb\ generates a \mev, taking into account the congestion history, to ensure packets are sprayed across congestion-free paths. To this extent, when a receiver node receives a packet, it copies the \ev\ field from the packet into the \ack\ and sends it back to the sender. Upon receiving an \ack, the sender checks it for congestion signals, \eg\ \ecn, and updates the congestion history accordingly. Intuitively, if an \ack\ carries an \ecn\ signal, \pslb\ considers a penalty (\ie\ $P_{\ecn}$) for the path associated with the \ev, attempting to avoid that path for some time. 
Similarly, when a sender receives a \nack\ packet, it marks the \ev\ carried by the \nack\ as congested. However, since a \nack\ indicates severe congestion, \pslb\ reflects this severity in the congestion history (\ie\ the penalty of a path associated with such an \ev\ is higher than that of associated with an \ecn, \ie\ $P_{\nack}>> P_{\ecn}$).

When sending a new packet, the sender generates a \mev\ (see Subsection~\ref{design_elements}) and, using the congestion history, ensures that the entire path is congestion-free, \ie\ none of the \ev s included in the \mev\ is experiencing congestion. If congestion history indicates that any \ev\ in \mev\ is experiencing congestion, the host skips that \ev\ and generates a new \ev. If all possible paths are congested, \pslb\ chooses the \ev\ with smallest penalty. After each \mev\ generation, the host decays the penalty values in the congestion history. The decaying mechanism is added to avoid stale congestion history information and guarantees that a path, appearing to be congested, will be eventually selected for packet spraying.  

Alg.~\ref{sender_host} shows the operation performed on the sender host.

\begin{algorithm}[!h]
	\caption{\pslb\ operation at sender host}
	\label{sender_host}
	\SetKwProg{Fn}{Function}{:}{}
	\Fn{Receive(packet)}{
	\tcc{Receive \ack\ or \nack\ by the sender}
		\eIf{packet == \nack}{
			CongestionHistory[\ev] = $P_{\nack}$;
			}{
			\If{packet==\ack\ and \ecn \ and CongestionHistory[\ev]==0}{
				CongestionHistory[\ev]=$P_{\ecn}$;
			}
			
			}
		}
		
		\Fn{onSend(packet)}{
		\eIf{CongestionHistory == NULL}{
			\tcc{First bdp\\
				No congestion history available}
			\mev.generate()\;
			Break;
		}{
			\tcc{Congestion history can be used;}
			\ev= \mev.generate()\;
			\For{\ev\ in \mev}{
				\While{CongestionHistory[\ev]!=0}{
					\tcc{Ensure that no \ev\ part is experiencing congestion}
					\ev= \mev.generate();
				}
			}
			\tcc{Decay congestion penalty of penalized paths.}
			CongestionHistory.decay()\; 
			\tcc{send the packet}
		}
	}

\end{algorithm}

\begin{figure}[!t]
	\centering
	\includegraphics[width=0.8\columnwidth]{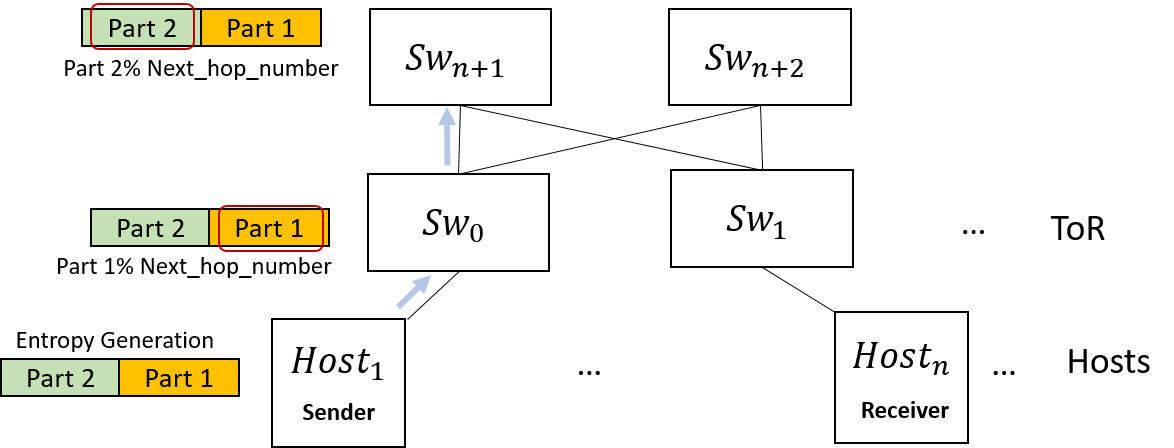}
	\caption{Switch process in \pslb.}
	\label{swicth_process}
\end{figure}

\cat{Process at switches} When a packet is received by a switch, the switch extracts the \ev\ from the packet's header field and uses it to forward the packet to the next hop. Recall that \pslb\ uses \mev; as such, the switch must extract the part in the \mev\ that corresponds to its tier and uses that part for packet forwarding. This process is highlighted in Fig.~\ref{swicth_process}.

\subsection{Design Elements}
\label{design_elements}
\cat{Multi-part entropy (\mev)} The \pslb's entropy field in the packet header is divided into multiple parts based on the number of tiers in the topology.
For instance the entropy field in a $3$-tier FatTree topology consists of two parts (see Fig.~\ref{swicth_process}). The value of each part specifies the index of the next-hop switch uplink port. As such, the number of bits allocated to each part is based on the number of uplink ports on switches within that tier of the topology. For example, if switches in the first tier (\ie\ tier~$1$) have $16$ uplink ports, $4$ bits are required for part~$1$ of the entropy. 

\cat{Generating \mev} When a packet is ready for transmission, an \ev\ for each part of the \mev\ is generated, concatenated and placed in the entropy field of the packet header. Switches at each tier are configured to extract and utilize the correct part (\ie\ \ev) from the \mev, based on the \ev\ length used for each tier. The main idea behind such design is that balancing the traffic in the uplinks, prevent downlinks to be overloaded in Section~\ref{theory} (Theorem~\ref{theorem1}).

\begin{figure}[!t]
	\centering
	\includegraphics[width=0.9\columnwidth]{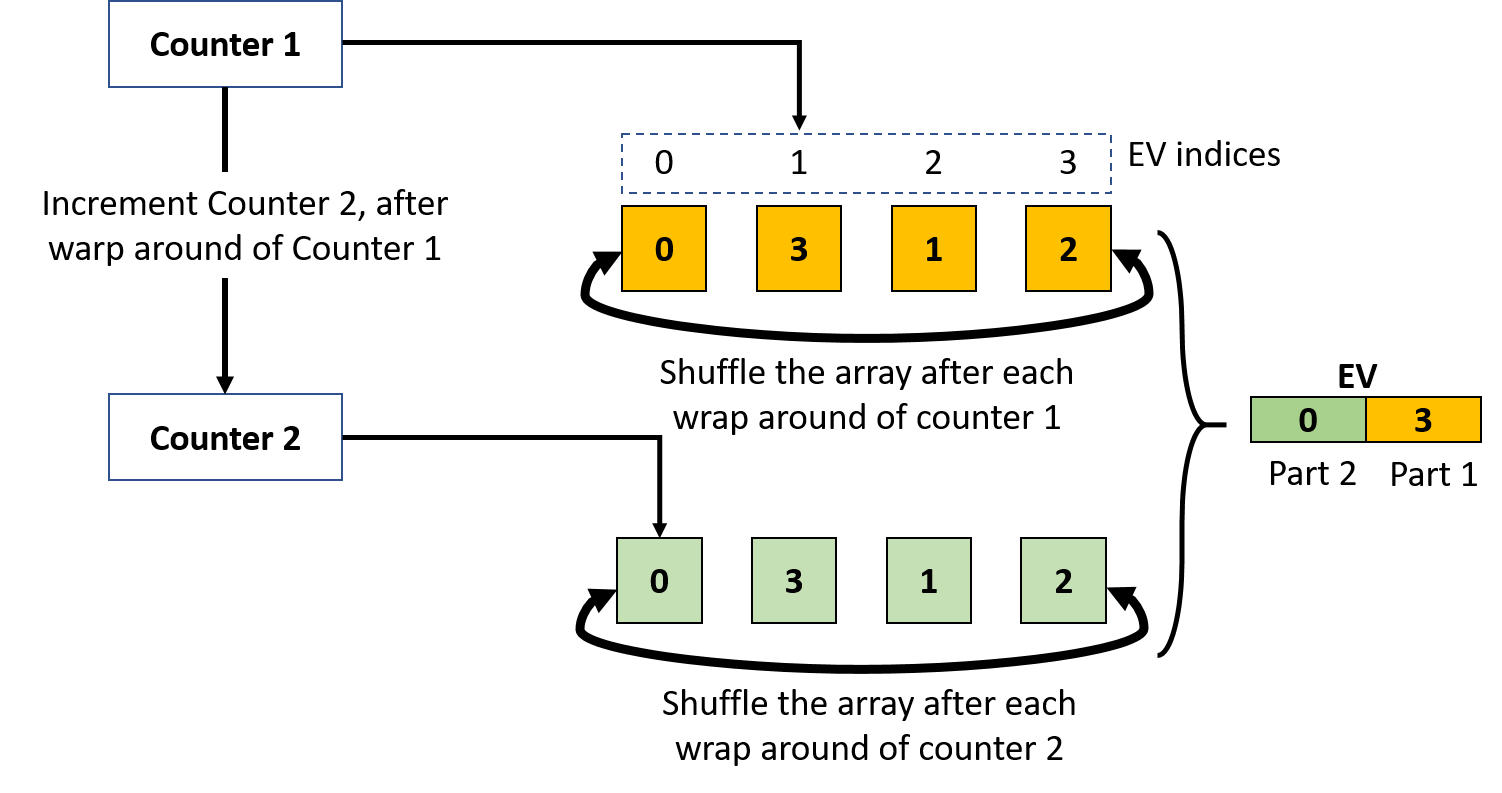}
	\caption{Example of generating two-part \ev.}
	\label{multipart-entropy}
\end{figure}

We generate each part in the \mev\ using a permutation-based method that follows a pseudo-random round-robin selection pattern (see Alg.~\ref{permutation_algorithm}). In particular, each \ev\ part in the \mev\ has an associated array whose length corresponds to the maximum number of upstream switch ports in a network tier (see line~\ref{evpart}). Each array stores a permutation of the upstream port indices. Moreover, each array is paired with a counter used to iterate over the array in a round-robin manner. This means that the counter iterates over the indices of the array (see line~\ref{counter}), and the value stored at each entry corresponds to the selected port (see line~\ref{value}). Various strategies can be employed to increment counters whenever a new packet needs to be transmitted. We employed a dependent implementation, where the counter for the higher order part (\ie\ \ev\ for the higher tier of the network) is incremented once the counter for the immediate lower order is exhausted (see lines~\ref{start} to~\ref{end}). Fig.~\ref{multipart-entropy} shows an example of this procedure in a $3$-tier FatTree topology, where counter~$2$ is only incremented at the wraparound of counter~$1$. 
\begin{algorithm}[!h]
	\SetAlgoLined
	\KwResult{Generate a multi-part \ev\; }
	$part=TopologyTier -1$\;
	EntropyArray$=[[Shuffled port numbers]]*part$ \label{evpart}\;
	\tcc{Generate a counter for each \mev}
	$counter=[0]*part$\; 
	\mev=[]\; 
	rand= GenRand(0,100)\;
	index=Determine part corresponding to the value of the rand\;
	counter[index]+=1\label{counter}\;		
	value=EntropyArray[counter[index]]\label{value}\;
	
	\If {counter[index]==len($EntropyArray[counter]$)}{\label{start}
		\tcc{Shuffle values after each iteration}
		shuffle ($EntropyArray[counter]$)\;
		counter[index]=0\label{end}\;
	}
	\For{j in EntropyArray}{
		\If{j==index}{
			\mev.append(value)\;
		}{
			\mev.append(EntropyArray(j[counter[j]]))\;
		}
	}
	
	\Return {\mev}
	\caption{Multi-part \ev\ generation process in the host}
	\label{permutation_algorithm}
\end{algorithm}

Each permutation set is shuffled after all indices have been iterated through (\ie\ the counter wraps around). To reduce synchronization among flows originating from different hosts, it is recommended to use a random seed for the random generator employed in shuffling. We employ Fisher–Yates shuffle algorithm~\cite{fisheryates}, since it has a computational complexity of $\mathcal{O}(n)$ and performs in-place permutation, requiring no additional memory. Fig.~\ref{multipart-entropy} illustrates an example process of generating \mev\ in the host.

\cat{Congestion history and path penalization} Congestion history is a data structure maintained by a host to keep track of congestion status of paths and penalize the congested paths based on the severity of the congestion they are experiencing. While congestion history has been previously studied in the literature~\cite{le2024strack,luo2024seqbalance}, we present an optimized mechanism that efficiently tracks the congestion status across all paths. In particular, we keep the congestion history of $N$ available paths in an array, where the indices of the array specify the \ev, where each \ev\ uniquely represents a path. The value stored in an array is used for retaining penalty value for the path \ev\ represents. This column can be implemented using $k$ bits, where $k$ is a design parameter and represents the maximum penalty value). 

The following operation are performed to update the congestion history:
\begin{itemize}
	\item \textit{Congestion penalty}: The penalty values for all paths are initialized to zero, indicating all path are congestion-free. When a sender receives an \ecn-marked \ack\ or \nack\ packets, it penalizes the paths associated with the \ev s carried in the packet. In particular, when a sender receives an \ecn-marked \ack, \pslb\ penalizes the path if its penalty is $0$. However, if the path's penalty is not zero, \pslb\ avoids multi-penalization as it causes conservative path avoidance for an extended amount of time. When a sender receives a \nack\ the path's penalty is updated to the $P_{\nack}$. Note, since \nack\ indicates a more severe congestion status compared to \ecn-marked \ack, \pslb\ assigns a much higher penalty for a path associated with an \ev\ carried in a \nack packet, \ie\ $P_{\nack}>>P_{\ecn}$. Additionally, to avoid multi-penalization of the same path, \pslb\ avoids re-penalizing a path that is \ecn-marked.
	\item \textit{Decaying function}: \pslb\ updates the congestion history to avoid using stale information. Specifically, when sending a new packet all penalty values would be updated. The update value is calculated based on the drainage rate of the switch, which is close to the $P_{\ecn}$. This means that a path with $P_{\nack}$ would take longer time compared to a path with associated \ecn, to be reused a gain.
\end{itemize}
%
\section{Evaluation}

In this section, we used a production-level proprietary packet-based simulator to study the performance of \pslb, and compared it with alternative load balancing solutions.

\cat{Setup}
We implemented different sized FatTree topologies, specified in Table~\ref{simulation topologies}. In our setup, we assumed all links in the network to have the same bandwidth as this is a common design in production data centers. We evaluated three different bandwidth configurations of $100$~Gbps, $400$~Gbps and $800$~Gbps to analyze the impact of bandwidth on \pslb. Finally, all links have a delay of $600$~us.
\begin{table}[!t]
	\caption{Simulation topologies.}
	\label{simulation topologies}
	\footnotesize
	\centering
	\begin{tabular}{|c|c|c|}
		\hline
		\textbf{Tier number} & \textbf{Number of hosts} & S\textbf{witch port number} \\\hline \hline
		2           & 128             & 16                 \\ \hline
		2           & 2k              & 64                 \\ \hline
		2           & 8k              & 128                \\ \hline
		3           & 1k              & 16                 \\ \hline
	\end{tabular}
\end{table}

\cat{Traffic} We evaluated \pslb\ using various flow sizes of $2$~MB, $4$~MB, $8$~MB, $16$~MB and $32$~MB. In all our experiments the MTU were set to $4160$~B which is a common setup in AI data centers. Additionally, \ecn\ thresholds were set to $K_{min}=0.25$~BDP and $K_{max}=0.75$~BDP, where BDP is \textit{Bandwidth Delay Product} in the topology under study. In all experiments, trimming is enabled and packets get trimmed if the queue length exceeds one BDP. Finally, all experiments use \ack\ coalescing for every $4$ packets.

\cat{Algorithms} In addition to \pslb, we also
implemented the following algorithms for comparison:
\begin{enumerate}[label=\roman*.]
	\item \textit{Adaptive Routing (\ar)}: A network-based load balancing approach in which the switch selects the next hop based on its local information, \eg\ queue size.
	\item \textit{\reps}~\cite{smartttreps2024}: A hash-based packet-spraying scheme that uses positive feedback from the network to reuse congestion-free paths.
	\item \textit{\obps\ (\ob)}: A revised version of \pslb, in which congestion signals are ignored.
\end{enumerate}
\cat {Metrics} We used the following performance metrics to evaluate \pslb: 
\begin{itemize}
\item \textbf{Maximum Flow Completion Time (\fct)}: The longest time taken among all flows in a given workload or experiment, measured from when the first packet of a flow is sent to when the last packet of a flow is received.
\item \textbf{Ratio}: Represents a normalized measure of performance (\ie\ the maximum flow completion time (\fct)) relative to the maximum \fct\ in the ideal scenario. The closer the ratio to $1$ the closer the performance of the algorithm to the optimal.
\item \textbf{Queue Depth}: The highest number of packets observed in the switch queue over the experiment time.
\end{itemize}

\begin{figure*}[!t]
	\begin{subfigure}[t]{0.33\textwidth}
		\includegraphics[width=0.95\textwidth]{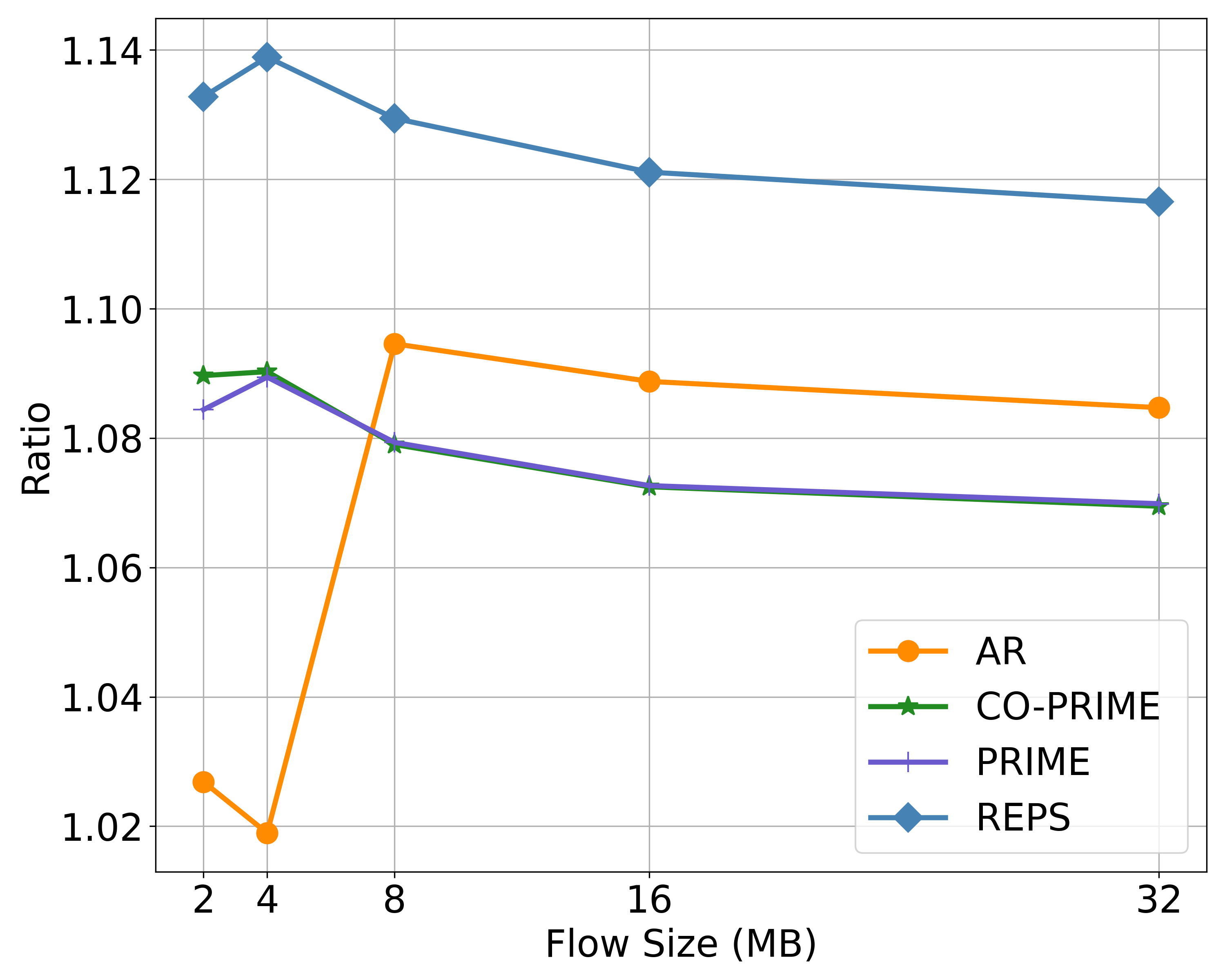}
		\caption{$100$~Gbps}
		\label{fig:permutation-100}
	\end{subfigure}%
	\hfill 
	\begin{subfigure}[t]{0.33\textwidth}
		\includegraphics[width=0.95\textwidth]{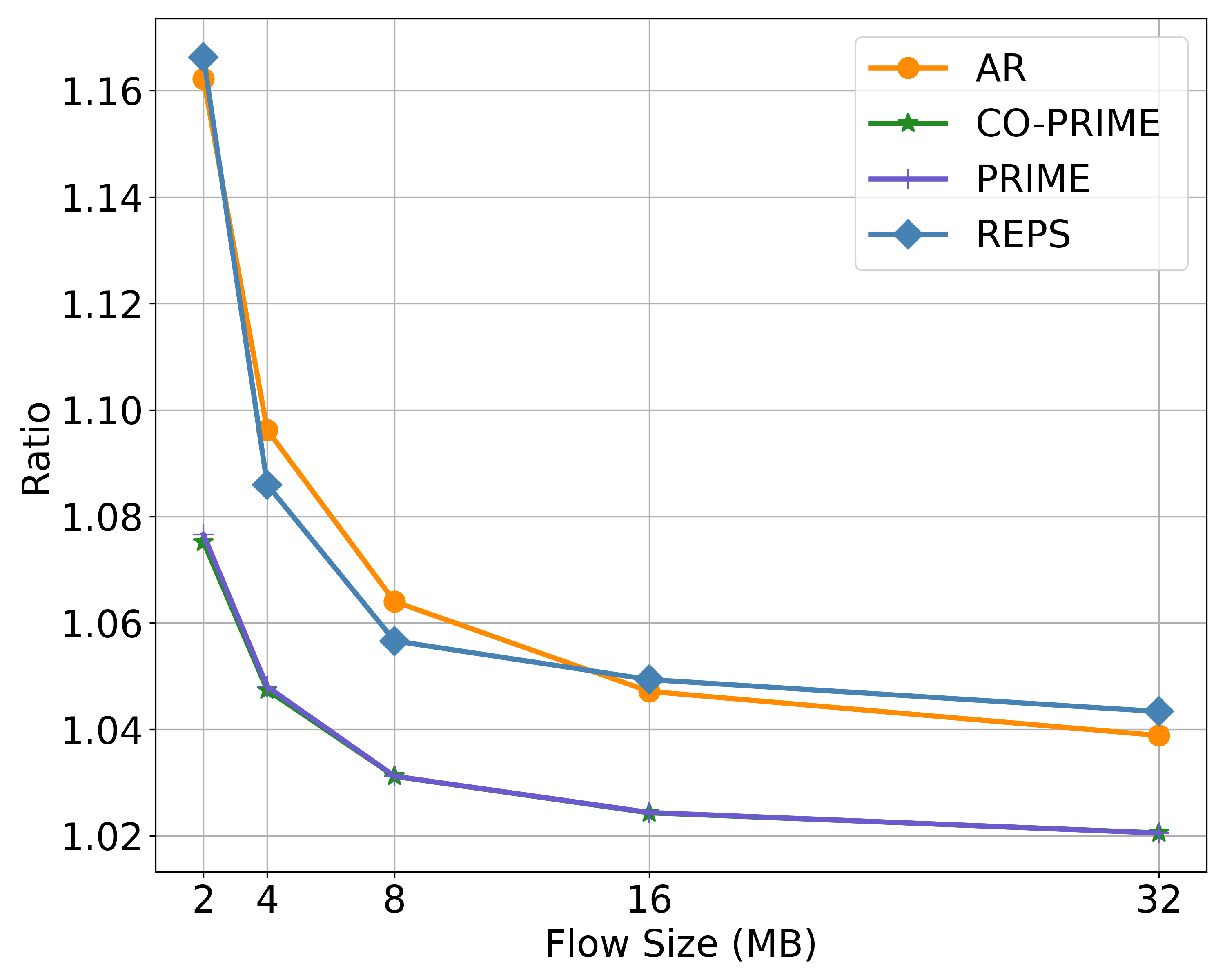}
		\caption{$400$~Gbps}
		\label{fig:permutation-400}
	\end{subfigure}%
	\hfill
	\begin{subfigure}[t]{0.33\textwidth}
		\includegraphics[width=0.95\textwidth]{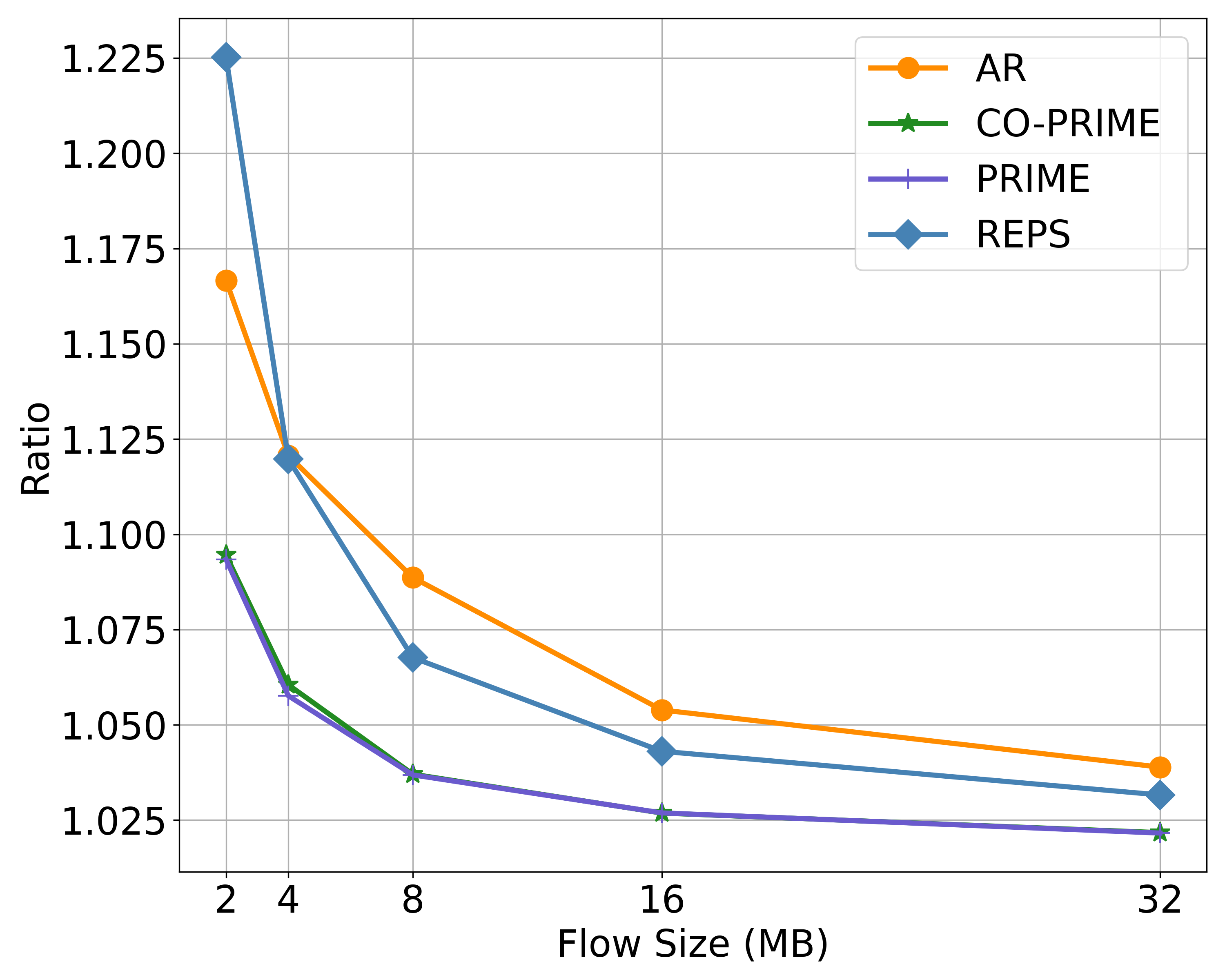}
		\caption{$800$~Gbps}
		\label{fig:permutation-800}
	\end{subfigure}%
	\caption{Permutation traffic in a $2$-tier FatTree topology with $2048$ hosts under varying link bandwidths.}
	\label{permutation-2tier}
\end{figure*}

\begin{figure*}
	\begin{subfigure}[t]{0.33\textwidth}
		\includegraphics[width=0.95\textwidth]{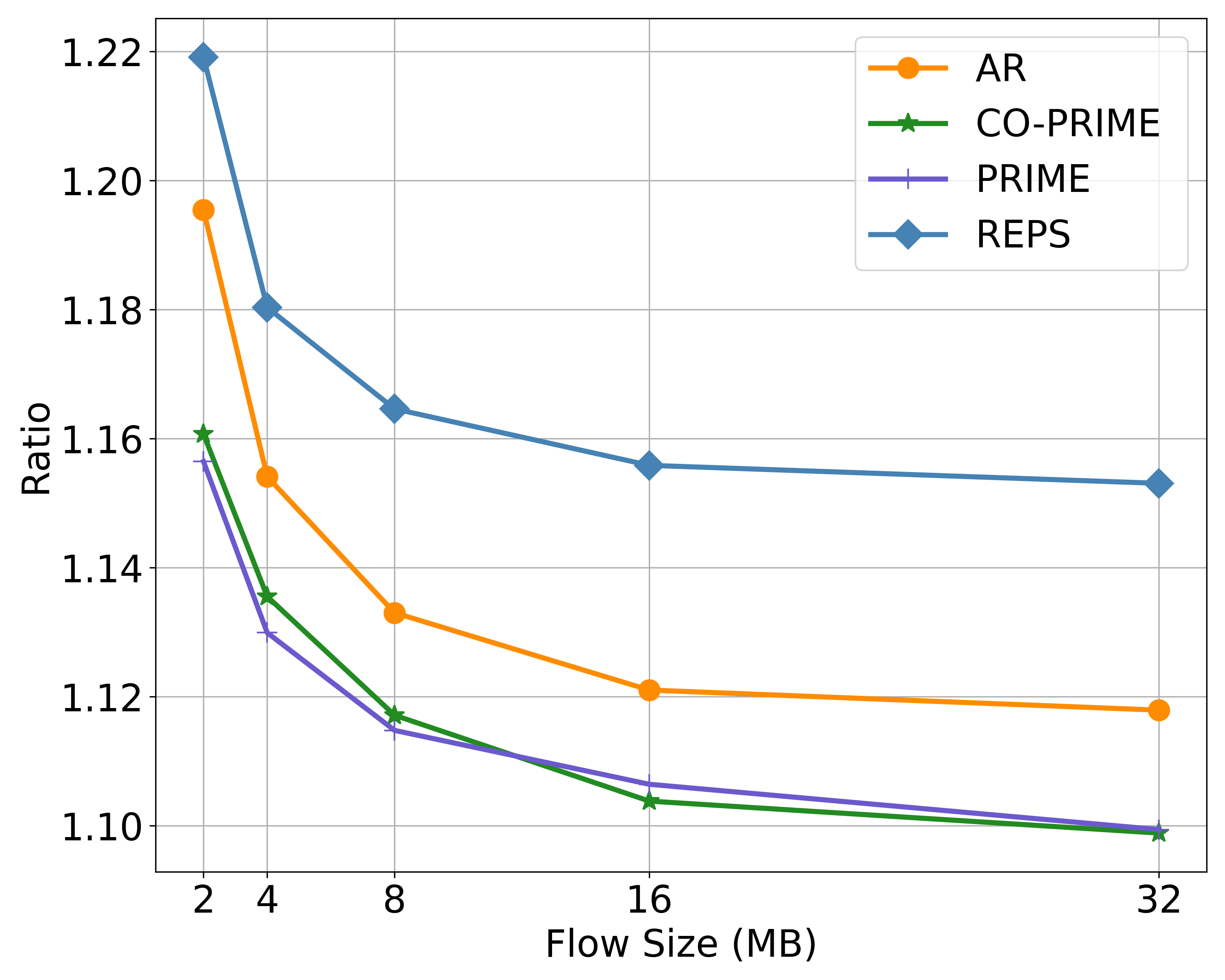}
		\caption{$100$~Gbps}
		\label{fig:permutation-100-3tier}
	\end{subfigure}%
	\hfill 
	\begin{subfigure}[t]{0.33\textwidth}
		\includegraphics[width=0.95\textwidth]{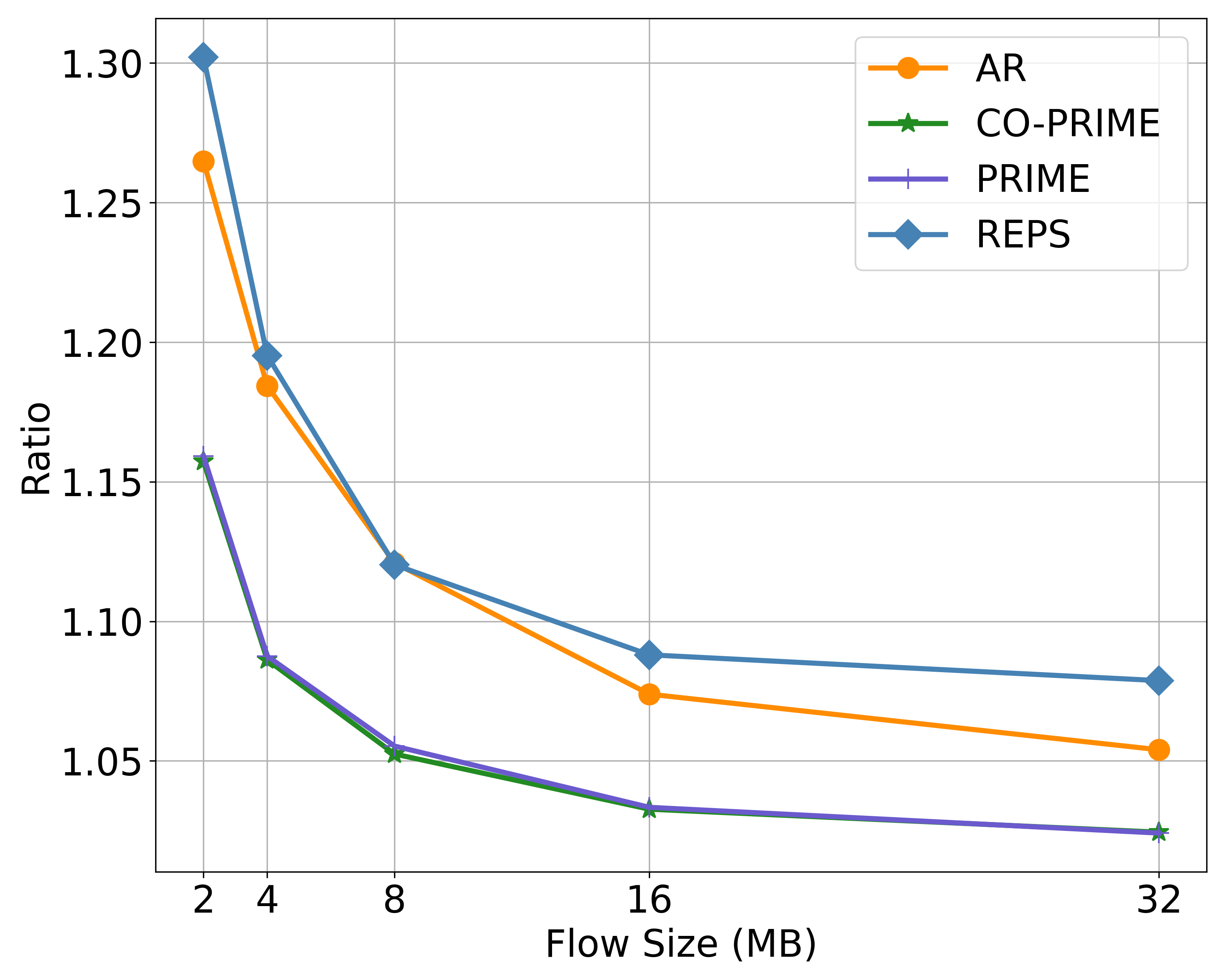}
		\caption{$400$~Gbps}
		\label{fig:permutation-400-3tier}
	\end{subfigure}%
	\hfill
	\begin{subfigure}[t]{0.33\textwidth}
		\includegraphics[width=0.95\textwidth]{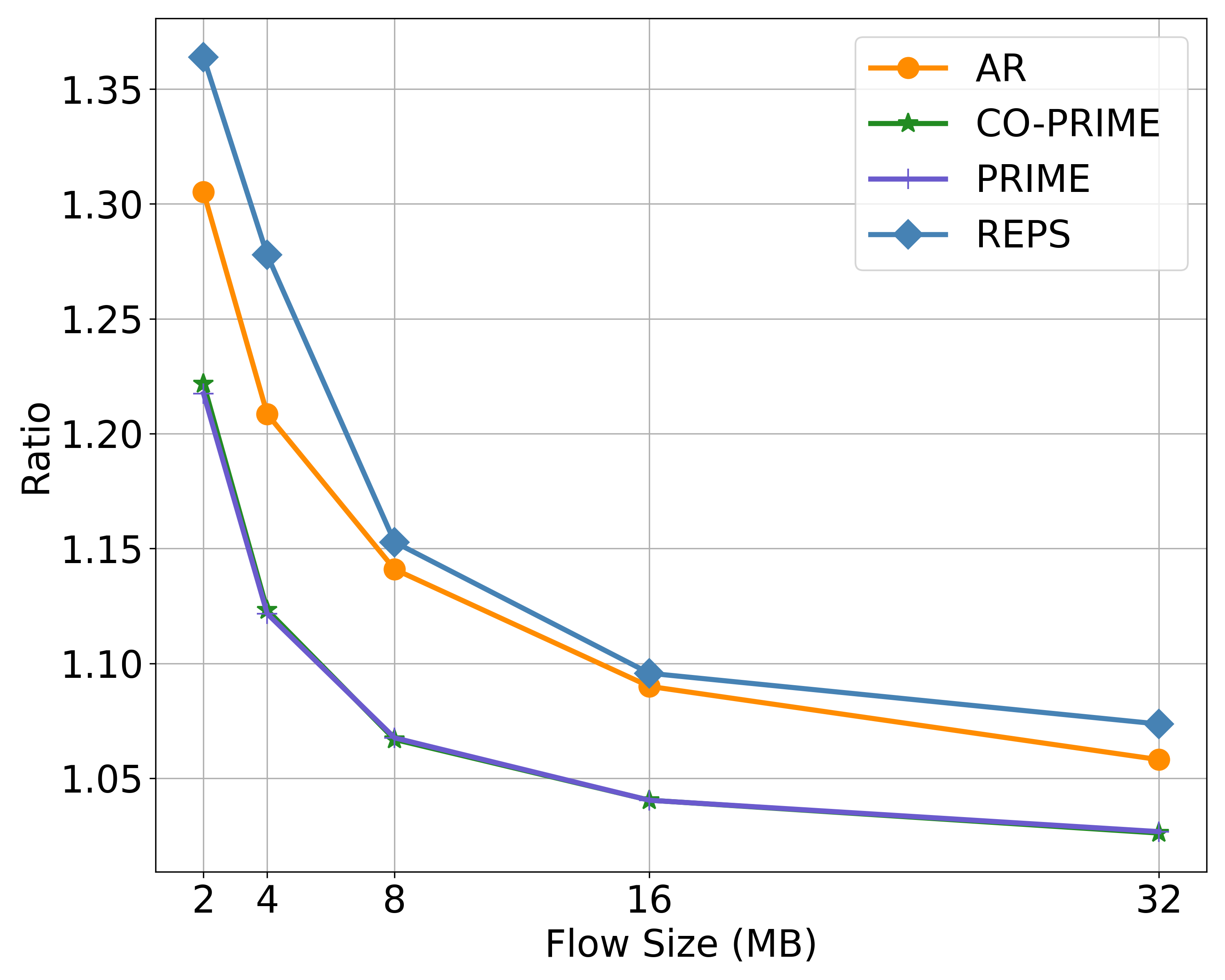}
		\caption{$800$~Gbps}
		\label{fig:permutation-800-3tier}
	\end{subfigure}%
	\caption{Permutation traffic in a $3$-tier FatTree topology with $1024$ hosts under varying link bandwidths.}
	\label{permutation-3tier}
\end{figure*}
\subsection{Results}
\subsubsection{Symmetric network conditions} The first
set of results focuses on a baseline healthy symmetric network condition.
\cat{Permutation traffic} In these experiments, we assessed the performance of \pslb\ against different load balancing algorithms in a scenario with perfectly symmetrical network with no over-subscription. Since this scenario involves no congestion, it offers an opportunity to evaluate how efficiently \pslb\ distributes packets across multiple links, unaffected by the congestion control, to achieve optimal performance.
Fig.~\ref{permutation-2tier} and Fig.~\ref{permutation-3tier} evaluate the performance of various load-balancing schemes under permutation traffic in $2$-tier ($2048$ hosts) and $3$-tier ($1024$ hosts) FatTree topologies, respectively, considering different link bandwidths and flow sizes. From figures, the following observations are in order. First, as bandwidth rises from $100$~Gbps to $800$~Gbps, smaller flows (flow size less than  $8$~MB) experience a more noticeable initial performance degradation (\ie\ higher Ratios), while for larger flows ($16$~MB and $32$~MB), all methods show improved performance, with Ratios close to optimal. This behavior occurs because shorter flows are heavily influenced by delays such as queuing and propagation delays, which are significant compared to their transmission time, resulting in a larger deviation from the ideal \fct. Larger flows, however, are primarily influenced by transmission time, bringing their \fct\ closer to the theoretical optimum.
Second, \pslb\ consistently outperforms other load-balancing schemes across both $2$-tier and $3$-tier FatTree topologies, under varying flow sizes and bandwidths, with a maximum performance improvement of up to $15\%$ compared to both \ar\ and \reps\ in a $3$-tier topology (for $2$~MB flow size at $800$~Gbps). This significant gain highlights \pslb's ability to mitigate non-transmission delay, \ie\ queuing delay, which is significant for smaller flows in high-bandwidth scenarios. The primary reason for the reduction in queuing delay is that \pslb\ distributes the traffic in a proactive round-robin fashion more efficiently than other approaches, leading to less buffer inflation and, consequently, a decrease in queuing delay.
Even for larger flows ($16$ and $32$~MB), where the performance of all methods tends to converge, \pslb\ consistently maintains its advantage, delivering a performance that is only less than $5\%$ below the ideal performance for at $400$~Gbps and $800$~Gbps scenarios. 

Third, all methods benefit from increased bandwidth, particularly in $3$-tier FatTree topologies. This occurs because, in permutation traffic, some flows bypass the spine switches. Furthermore, higher link speeds increase the available bandwidth for flows that traverse the spine switches, thereby reducing the chances of congestion and queue buildup. In other words, with more capacity available in the network, the impact of load balancing decisions and potential imbalances becomes less severe, leading to improved performance for all load-balancing schemes. 

Finally, a comparison between \pslb\ and \ob\ reveals that their results remain consistently similar across various scenarios. The reason is that in this scenario no congestion exists in the network. As such, adaptive consideration of congestion signals does not improve the performance of the load balancing scheme. As such, \ob\ which is non-adaptive and does not consider congested paths in the network achieves similar performance to \pslb. Therefore, packet spraying technique used in both methods is highly efficient in minimizing buffer buildup.

\begin{figure}[!h]
	\centering
	\includegraphics[width=0.7\columnwidth]{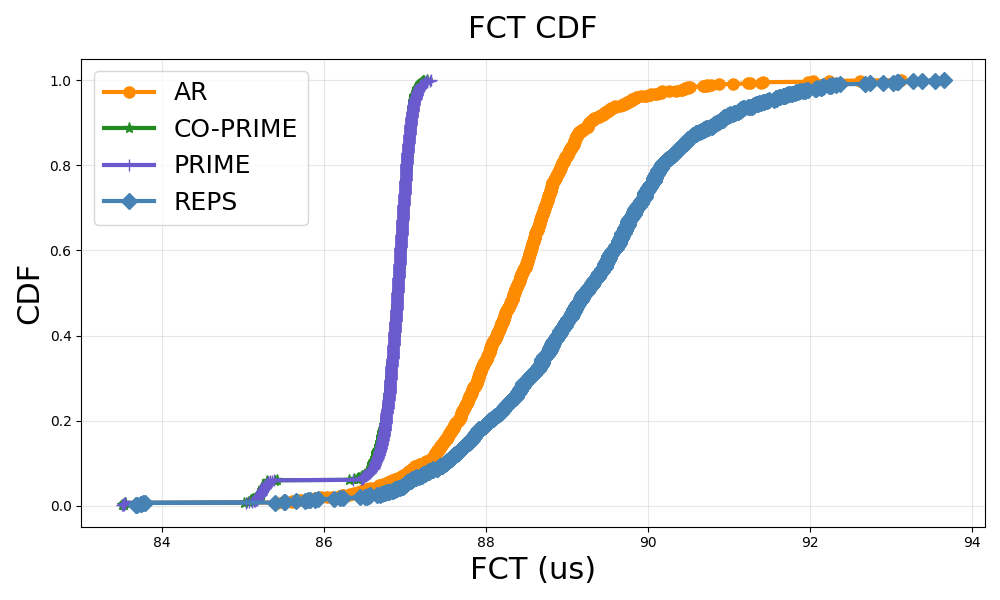}
	\caption{Average \fct\ for $8$~MB flows in $3$-tier FatTree topology with $800$~Gbps.}
	\label{fig:avg_fct}
\end{figure}

While tail latency is vital for optimizing the performance of AI workloads, the \fct\ remains significant, particularly in scenarios where the flows are associated with distinct jobs. Fig.~\ref{fig:avg_fct} depicts the performance of \pslb\ in such scenario in a $3$-tier FatTree topology with $800$~Gbps links and flow sizes of $8$~MB. The figure illustrates that \pslb\ not only improves the tail latency, but also enhances the average \fct\ compared to both \reps\ and \ar. This is particularly important in environments where multiple AI jobs run concurrently on the same AI fabric, as it ensures fair communication across jobs. While tail latency is critical for a single AI job, fairness across the tail latencies of different jobs is essential, given that the AI fabric does not inherently distinguish between them. Ensuring fairness among different flows prevents job starvation and balances network utilization, ultimately improving overall system efficiency.

\begin{figure}[!h]
	\begin{subfigure}[t]{0.489\columnwidth}
		\includegraphics[width=\columnwidth]{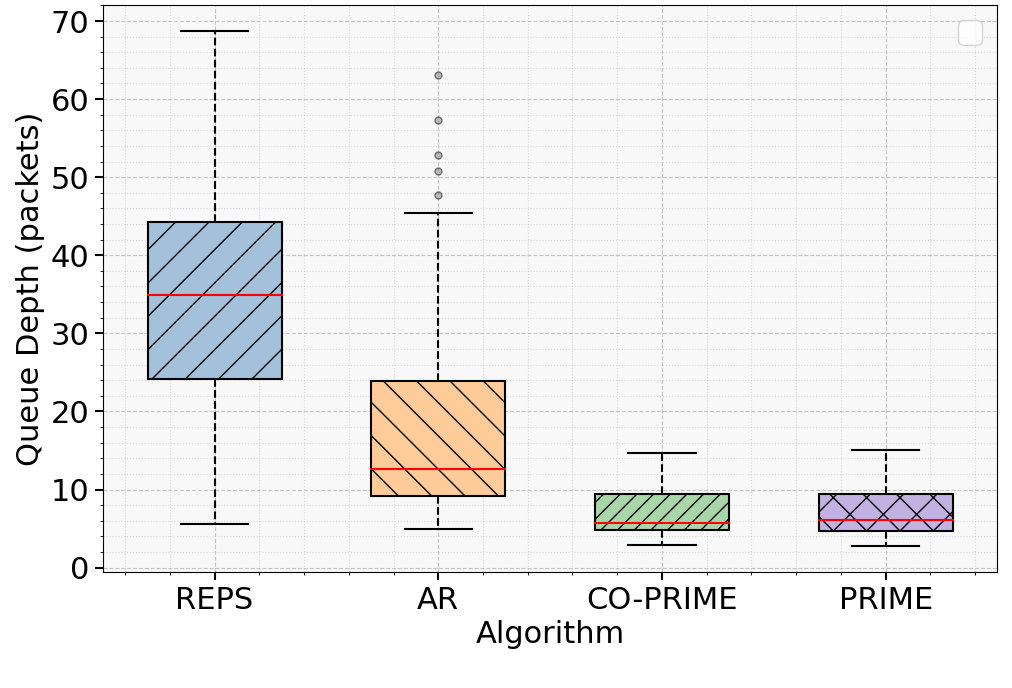}
		\caption{Buffer occupancy distribution.}
		\label{fig:buffer-3tier-box}
	\end{subfigure}%
	\hfill 
	\begin{subfigure}[t]{0.45\columnwidth}
		\includegraphics[width=\columnwidth]{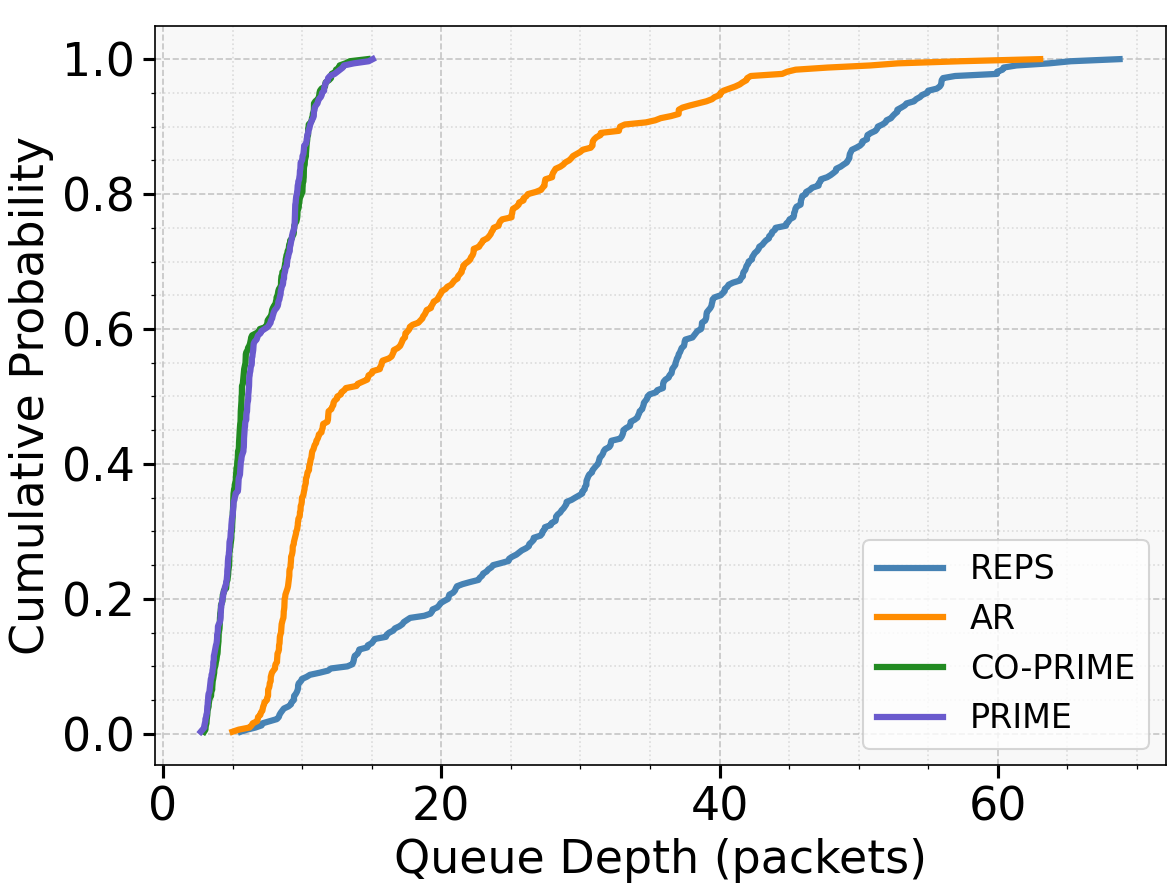}
		\caption{Average buffer occupancy.}
		\label{fig:buffer-3tier-average}
	\end{subfigure}%
	\caption{Buffer occupancy for $8$~MB flows in a $3$-tier FatTree topology with $800$~Gbps links.}
	\label{buffer-3tier}
\end{figure}

\cat{Buffer analysis}
To further explain why \pslb\ achieves a lower Max-FCT compared to \reps\ and \ar, Fig.~\ref{buffer-3tier} depicts the buffer occupancy recorded for different load balancing algorithms for $8$~MB flows in a $3$-tier FatTree topology with $800$~Gbps links. The Fig.~\ref{fig:buffer-3tier-box} shows the buffer occupancy distribution, where \pslb\ exhibits significantly lower queue depths compared to \reps\ and \ar. The Fig.~\ref{fig:buffer-3tier-average} presents the distribution of the average buffer occupancy, further highlighting that \reps\ maintains a consistently lower queue depth. This reduction in buffer occupancy, both in terms of average and deviation, plays a crucial role in minimizing queuing delays, ultimately contributing to improved flow completion times. In contrast, \reps\ and \ar\ experience higher buffer occupancy, leading to increased queuing delays and, consequently, higher flow completion times. The ability of \pslb\ to spray packets more efficiently and to prevent excessive queue buildup, results in superior performance, especially for smaller flow sizes where queuing delay is comparable to transmission delay.

\subsubsection{Asymmetric network conditions} We evaluate the
performance of \pslb\ under asymmetric network conditions, including \textit{link failure}, \textit{Link degradation} and \textit{coexistence of packet sprayed traffic with non-sprayed traffic}.

\cat{Link failure} 
In data centers, link failures are inevitable, and recovery times can vary, typically ranging up to several milliseconds. The recovery speed depends on the detection and failover mechanisms in place, such as Bidirectional Forwarding Detection (BFD)~\cite{BFD}. To assess load balancing algorithms in this scenario, the process is divided into two stages: \begin{enumerate*}\item transient phase: occurs before failed links are restored, where packets are sent to the failed links, thereby resulting in a blackhole;
\item  steady phase: happens after recovery, where failed links are excluded from forwarding paths and omitted in load balancing.
\end{enumerate*}  Modern data center switches feature advanced capabilities (\eg\ DPFR~\cite{dpfr} and Cognitive Routing~\cite{broadcom}) that enable detection and recovery from link failures in sub-millisecond timeframes, suggesting that the transient phase is brief enough not to significantly affect the job completion times. However, in the steady phase, removing failed links from data forwarding creates an imbalanced topology with reduced bandwidth. In such cases, congestion control should adjust traffic rates to align with the new available bandwidth, though this adjustment takes time. During this period, an optimal load balancing algorithm should distribute traffic evenly so that all flows are equally affected by the bandwidth reduction, thereby minimizing tail latency. 

\begin{figure}[!h]
	\centering
	\includegraphics[width=0.3\textwidth]{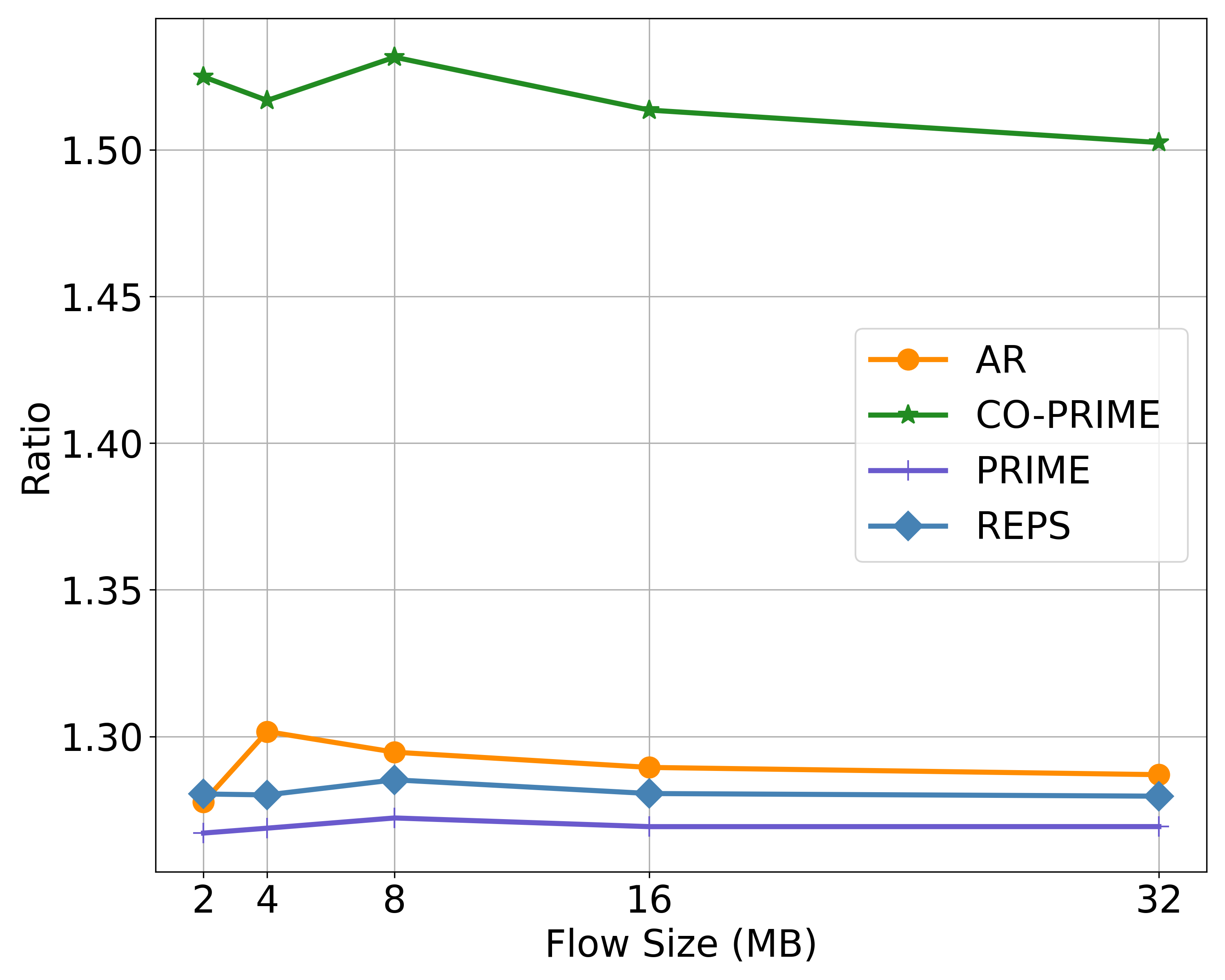}
	\caption{Link failure.}
	\label{fig:link_failure}
\end{figure}

To evaluate \pslb\ in such imbalanced scenarios, an experiment is designed where two links are failed in a $2$-tier topology with $128$~nodes. As depicted in Fig.~\ref{fig:link_failure}, the results reveal that \pslb\ consistently achieves the lowest ratio, indicating better performance in handling link failures with minimal impact on network efficiency. As anticipated, \ob\ conversely displays the highest ratio (around $20\%$ higher ratio compared to \pslb), since it fails to consider the temporary congestion in the network resulting from the imbalanced load caused by link failures.

\begin{figure}[!h]
	\centering
	\includegraphics[width=0.3\textwidth]{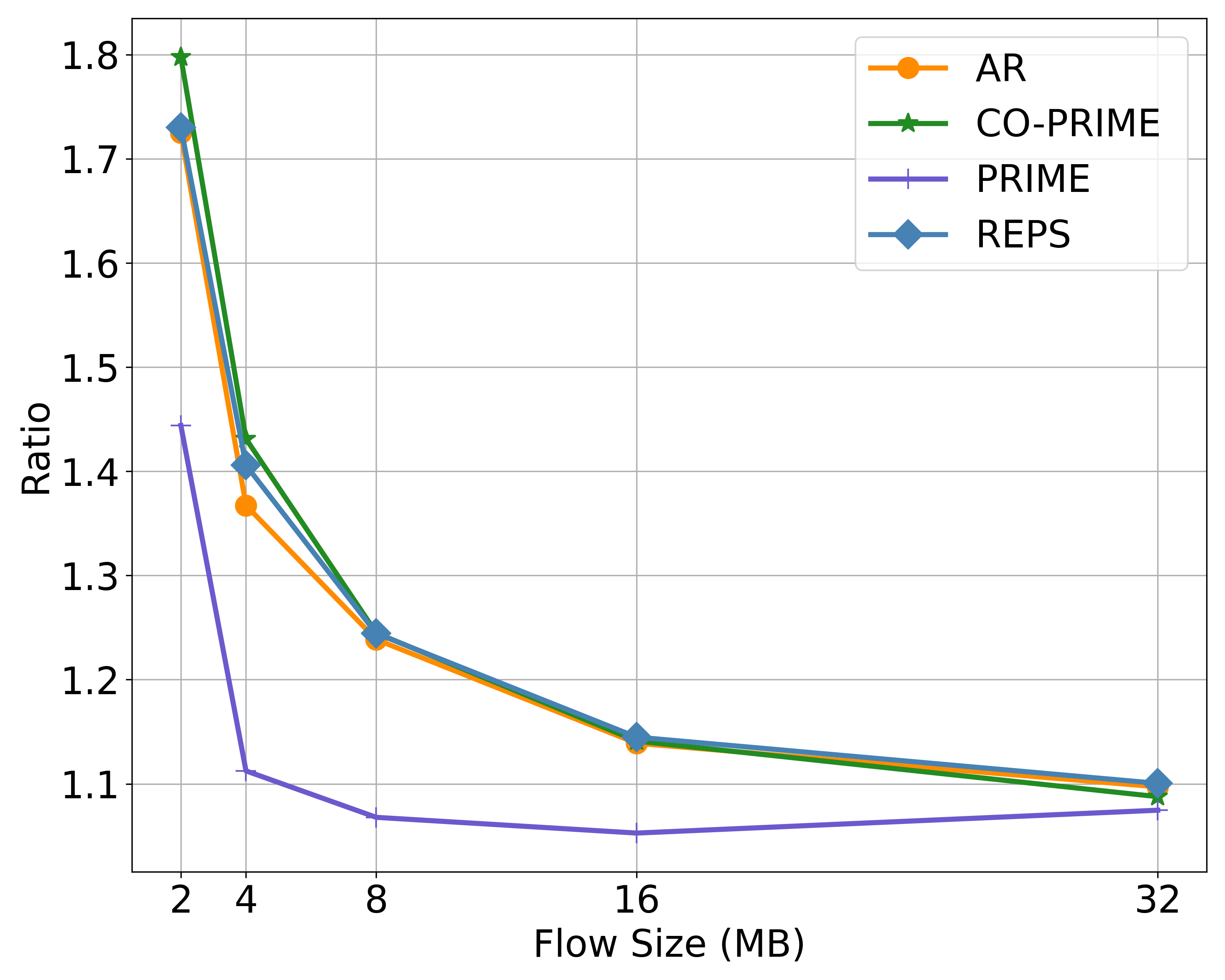}
	\caption{Link degradation.}
	\label{fig:link_degradation}
\end{figure}

\cat{Link degradation}  
In this experiment, we simulate link degradation, which could occur due to multiple factors such as: SerDes degradation, coexistence of strict-ordered traffic (\eg\ In-Network Computing traffic) and unordered traffic, hardware failures, or environmental interference. In symmetric topologies, distributing traffic equally across all paths, irrespective of traffic conditions, ensures optimal load balancing. However, this is not the case in asymmetric topologies, where there is one or more links degraded. Our goal is to evaluate \pslb\ in such scenarios. To this extent, we simulate a network with $8$~k hosts and the link speed of $400$~Gbps. $25\%$ of the links are randomly degraded to $100$~Gbps, representing the existence of in-network communication (INC) traffic. In particular, $75\%$ of the link bandwidth on the chosen links are allocated to INC, further exacerbating the challenges of load balancing in such environments. Fig.~\ref{fig:link_degradation} illustrates the impact of link degradation on the performance of different load balancing strategies in a network. The findings demonstrate that \pslb\ consistently outperforms other strategies across all flow sizes,showing a notable advantage with smaller flows (around $27\%$ improvement compared to \reps\ for $4$~MB flows), whereas \ar, \reps\ and \ob\ exhibit comparable, moderate performance. Notably, despite \ob’s inability to respond to network congestion, it still delivers results comparable to \ar\ and \reps. This highlights that by distributing traffic across multiple paths, even without adaptive adjustments, the effects of degraded links are dispersed among all affected flows, reducing the impact on each individual flow. Thus, traffic spraying, even in a non-adaptive manner, can offer a degree of resilience in such scenarios.

\cat{Mixed flow (coexistence of packet sprayed with non-sprayed traffic)} In these experiments, the performance of \pslb\ and \reps\ are evaluated within a $3$-tier FatTree topology, under conditions where both \ecmp\ traffic (\ie\ non-sprayed) and packet sprayed traffic coexist. In particular, the \ecmp\ traffic constitutes $1\%$ of the total $1024$ flows (\ie\ around $10$ flows). We compare the Maximum \fct\ values of \pslb\ and \reps\ under various scenarios, where different packet scheduling algorithms such as strict priority (\spa) and Weighted Round Robin (WRR), is used for packet forwarding. In particular, when using \spa, \ecmp\ traffic has higher priority than packet sprayed traffic, while in WRR, different weights are allocated to \ecmp\ and packet sprayed traffic. We assigned \ecmp\ traffic weights of $50\%$, $66\%$, and $80\%$. Finally, to isolate the effects of load balancing from those of congestion control, we use the same congestion control mechanism for both \ecmp\ and packet-sprayed flows. 

\begin{figure}[!t]
	\centering
	\includegraphics[width=\columnwidth]{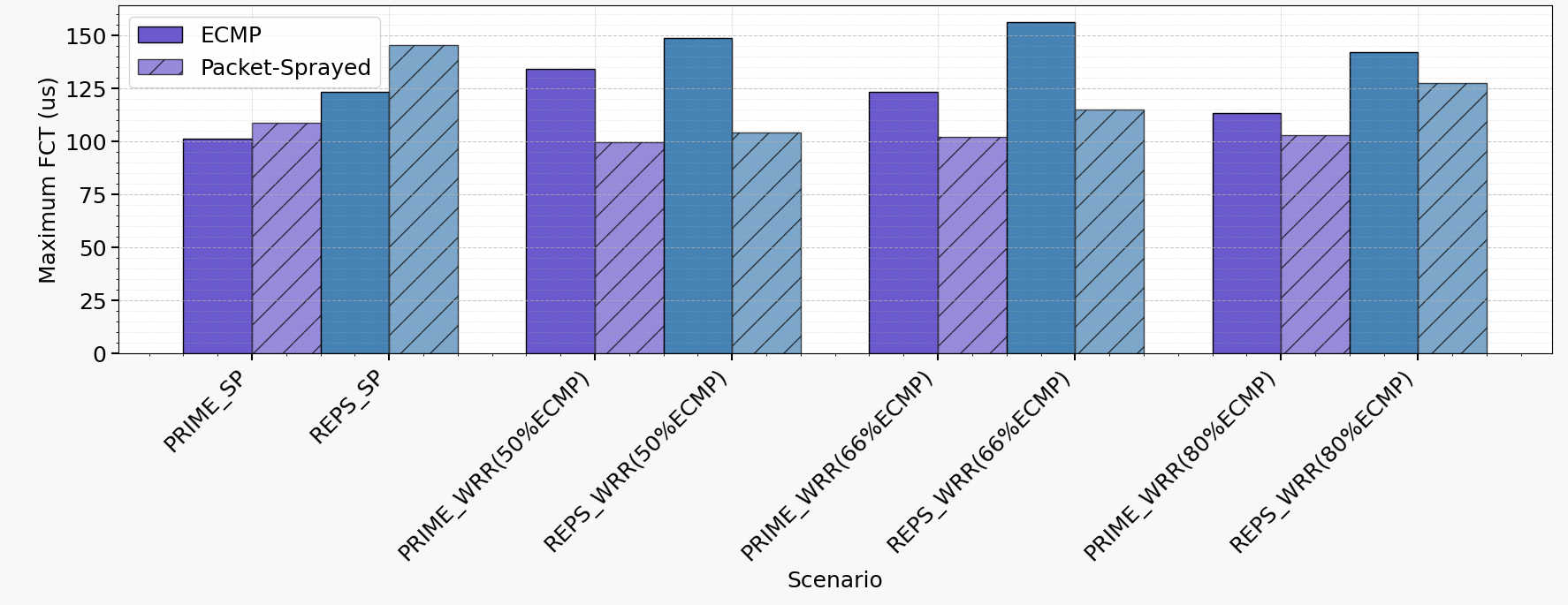}
	\caption{Mixed flow.}
	\label{fig:mixed_flow}
\end{figure}

Fig.~\ref{fig:mixed_flow} illustrates the comparison results. The following observations are in order. First, for both algorithms (\ie\ \reps\ and \pslb), \ecmp\ traffic has its lowest maximum \fct, when strict priority is used. The effect on \ecmp\ maximum \fct\ is due to buffer build up in the network, resulting in trimmed packets and degraded performance. In this case, the sprayed traffic packets can be starved by the \ecmp\ traffic packets, resulting in time-out and re-transmission, which affects their \fct. 
Second, when using WRR, as the allocated weight to \ecmp\ traffic increases—eventually resulting in strict prioritization and \ecmp\ traffic preempting non-sprayed traffic—the performance of sprayed traffic under \pslb\ is not impacted much. However, sprayed traffic in \reps\ are largely affected. This suggests that \pslb\ effectively avoids paths being used by \ecmp\ traffic under stricter scheduling policies. 
Third, Under WRR scheduling with equal bandwidth allocation, \ie\ $50\%$ \ecmp, \ecmp\ flows experience the most significant performance degradation, due to collision between \ecmp\ and sprayed-traffic. Since the sprayed packets belong to different flows, the impact of any congestion signal is lower on these flows.
Third, across all scheduling scenarios, when \pslb\ is used, the impact on both sprayed and \ecmp\ traffic is notably reduced compared to when \reps\ is used. The reason is that \pslb\ distributes packets more evenly across available paths compared to \reps, ensuring that all sprayed flows are equally exposed to congestion. 

\section{Related Works}
In this section we categorize and cover solutions that were used for load balancing in lossy networks. 

\cat{=-level} \ecmp, while being widely adopted in data centers, suffers from inefficiencies in handling traffic imbalances, particularly large, long-lived flows that can lead to congestion. To address \ecmp's limitations, Hedera~\cite{259355} and MicroTE~\cite{benson2011microte} proposed centralized flow scheduling solutions for multi-tier networks to dynamically balance elephant flows across different paths. Unlike MicroTE which proactively routed traffic based on short-term demand predictions, Hedera reacted to real-time congestion by dynamically rerouting elephant flows using network utilization data. Mahout~\cite{curtis2011mahout} and Planck~\cite{rasley2014planck} reduced the scheduling overhead and time of these solutions, respectively. In particular, Mahout, detected and marked elephant flows and signaled the controller to re-route and balance these flows, while Planck~\cite{rasley2014planck} sent small probes to measure congestion status in the network. These solutions, however, are centralized approaches and had long control loops. To omit dependence on a centralized controller, FlowBlender~\cite{kabbani2014flowbender} proposed an end-host-driven load balancing approach to assign flows to different paths.
Flow-level approaches, however, are coarse-grained
due to the large time constraints of their control loops and as such cannot be used ia AI/ML data centers due to their short latency requirement.

\cat{Sub-flow and Flowlet-level} To enhance path diversity and react faster to network conditions, some approaches, such as~\cite{alizadeh2014conga,katta2016hula,katta2016clove,vanini2017let,he2015presto} split flows into smaller bursts, \ie\ flowlets, and dynamically distribute flowlets in the network. Conga~\cite{alizadeh2014conga} used link utilization information to distribute flowlets across different paths. Conga, however, requires per-path congestion state on the leaf switches, and as such, is not scalable. Hula~\cite{katta2016hula} reduces the memory requirement of switches in Conga by requiring each switch to decide on the next hop of a flowlet. LetitFlow~\cite{vanini2017let} further reduces the complexity by deciding on the path for each flowlet randomly. These solutions, however, use a predefined time gap to define flowlet, which greatly affects their performance. 
Presto~\cite{he2015presto} performed load balancing at the edge by splitting flows into fixed size flow cells and assigning them to paths using hashing. Sub-flow and flowlet-based load balancing schemes, however, are challenging to be adopted in today's data centers due to using fixed sized idle gaps. fixed size  high-speed links, low latency requirements and asymmetric congestion.

\cat{Packet-level}
Packet-level load balancing distributes individual packets of a flow across multiple paths. 
FastPass~\cite{perry2014fastpass} is a load balancing scheme, where a central controller dynamically selects the transmission path and time for each packet based on factors such as flow size and destination. Centralized schemes, however, are impractical for AI/ML data centers due to their high latency and limited scalability. As such, distributed data-plane load balancing schemes, such the works~\cite{ghorbani2017drill, dixit2013impact,cao2013per,smartttreps2024} are preferred to provide scalability and adaptability to dynamic traffic patterns. In particular, Drill~\cite{ghorbani2017drill} employs per-packet decisions at each switch based on local queue occupancy. Drill, however, requires continuous queue monitoring, and as such, introduces complexity at switches. To reduce the complexity,
RPS~\cite{dixit2013impact} distributes packets randomly among the available shortest paths. RPS, however, does not account for the network's current state and, therefore, cannot prevent hotspots. To account for the network state, \reps~\cite{smartttreps2024} uses positive feedback from the network to reuse non-congested paths. This approach, however, results in stale information when \ack\ coalescing is used. Additionally, both RPS and \reps\ result in queue build-up and large network latency due to utilizing hash-based \ev s. To address the above limitation, the work~\cite{cao2013per}, selects packet's path using a deterministic Round-Robin (\rr) algorithm. Orthogonal to the previously mentioned works, QDAPS\cite{huang2018qdaps} and AG~\cite{liu2019ag} alleviated the packet reordering. While the proposed approach distributes traffic evenly, it ignores congestion information, leading to suboptimal performance.
%

\section{Discussion}
\cat{Other data center topologies} In this paper, we have focused on FatTree/Clos topologies as they are the most common topologies used in AI/ML data centers. However, \pslb\ can be extended to account for other topologies, \eg\ dragonfly. The main step is to consider the mapping between each part iof \mev\ to the topology.

\cat{Packet re-ordering} Multipathing can lead to out-of-order delivery at NICs; even when data packets are directly DMAed to hosts, memory is still required to track packet arrival information. \pslb, however, causes minimal re-ordering. However, \pslb\ results in minimal reordering because multipathing only introduces reordering when the delay difference between paths exceeds the inter-packet gap within a flow. Queuing delay is the dominant source of network delay in data centers and \pslb's low variance among queue lengths indicates minimal reordering (see Fig.~\ref{fig:buffer-3tier-box}).
\section{Conclusion}
In this paper, we presented the design and evaluation of \pslb, a pseudo-randomized round-robin approach to packet spraying for next generation data center networks tailored for AI/ML workloads. The key idea behind \pslb\ is to use multi-part entropy corresponding to the network topology and penalizes the paths experiencing congestion. To this extent, \pslb\ accounts for the congestion severity level and penalizes the paths based on the congestion level. Our extensive evaluation of \pslb\ in large-scale simulations show that \pslb\ constantly outperforms alternative packet spraying solutions, such as \reps, under various network configuration and different scenarios. In particular, our experiments indicate \pslb\ achieves up to $15\%$ for permutation traffic and up to $27\%$ improvement in asymmetric network conditions compared to \reps. As a worthwhile effort to extend this work, we believe it is possible to use more fine-grained congestion signals, \eg\ Congestion Signaling (CSIG)~\cite{csig} or in-band network telemetry~\cite{int}, into network conditions and avoid specific links. 

\clearpage
\bibliographystyle{IEEEtran}
\bibliography{./bib/main}

\begin{thebibliography}{10}
\providecommand{\url}[1]{#1}
\csname url@samestyle\endcsname
\providecommand{\newblock}{\relax}
\providecommand{\bibinfo}[2]{#2}
\providecommand{\BIBentrySTDinterwordspacing}{\spaceskip=0pt\relax}
\providecommand{\BIBentryALTinterwordstretchfactor}{4}
\providecommand{\BIBentryALTinterwordspacing}{\spaceskip=\fontdimen2\font plus
\BIBentryALTinterwordstretchfactor\fontdimen3\font minus
  \fontdimen4\font\relax}
\providecommand{\BIBforeignlanguage}[2]{{%
\expandafter\ifx\csname l@#1\endcsname\relax
\typeout{** WARNING: IEEEtran.bst: No hyphenation pattern has been}%
\typeout{** loaded for the language `#1'. Using the pattern for}%
\typeout{** the default language instead.}%
\else
\language=\csname l@#1\endcsname
\fi
#2}}
\providecommand{\BIBdecl}{\relax}
\BIBdecl

\bibitem{addanki2024challenging}
V.~Addanki, P.~Goyal, and I.~Marinos, ``Challenging the need for packet
  spraying in large-scale distributed training,'' \emph{arXiv preprint
  arXiv:2407.00550}, 2024.

\bibitem{lu2017flexflow}
W.~Lu, G.~Yan, J.~Li, S.~Gong, Y.~Han, and X.~Li, ``{Flexflow}: {A} flexible
  dataflow accelerator architecture for convolutional neural networks,'' in
  \emph{Proc. IEEE International Symposium on High Performance Computer
  Architecture ({HPCA})}, Feb 2017.

\bibitem{hoefler2022hammingmesh}
T.~Hoefler, T.~Bonato, D.~De~Sensi, S.~Di~Girolamo, S.~Li, M.~Heddes, J.~Belk,
  D.~Goel, M.~Castro, and S.~Scott, ``Hammingmesh: a network topology for
  large-scale deep learning,'' in \emph{Proc. International Conference for High
  Performance Computing, Networking, Storage and Analysis (SC)}, Feb 2022.

\bibitem{zhang2020network}
Z.~Zhang, C.~Chang, H.~Lin, Y.~Wang, R.~Arora, and X.~Jin, ``Is network the
  bottleneck of distributed training?'' in \emph{Proc. of the Workshop on
  Network Meets {AI \& ML} ({NetAI})}, Aug 2020.

\bibitem{alizadeh2014conga}
M.~Alizadeh, T.~Edsall, S.~Dharmapurikar, R.~Vaidyanathan, K.~Chu,
  A.~Fingerhut, V.~T. Lam, F.~Matus, R.~Pan, N.~Yadav \emph{et~al.}, ``{CONGA}:
  {Distributed} congestion-aware load balancing for datacenters,'' in
  \emph{Proc. in Special Interest Group on Data Communication ({SIGCOMM})}, Aug
  2014.

\bibitem{alizadeh2010data}
M.~Alizadeh, A.~Greenberg, D.~A. Maltz, J.~Padhye, P.~Patel, B.~Prabhakar,
  S.~Sengupta, and M.~Sridharan, ``Data center tcp ({dctcp}),'' in \emph{Proc.
  in Special Interest Group on Data Communication ({SIGCOMM})}, Aug 2010.

\bibitem{smartttreps2024}
T.~Bonato, A.~Kabbani, D.~De~Sensi, R.~Pan, Y.~Le, C.~Raiciu, M.~Handley,
  T.~Schneider, N.~Black, A.~Ghalayini, D.~Alves, M.~Papamichael, A.~Caulfield,
  and T.~Hoefler, ``{SMaRTT-REPS}: {Sender}-based marked rapidly-adapting
  trimmed and timed transport with recycled entropies,'' \emph{arXiv preprint
  arXiv:2407.00550}, 2024.

\bibitem{hoefler2023datacenterethernetrdmaissues}
\BIBentryALTinterwordspacing
T.~Hoefler, D.~Roweth, K.~Underwood, B.~Alverson, M.~Griswold, V.~Tabatabaee,
  M.~Kalkunte, S.~Anubolu, S.~Shen, A.~Kabbani, M.~McLaren, and S.~Scott,
  ``Datacenter {Ethernet} and {RDMA}: {Issues} at hyperscale,'' 2023. [Online].
  Available: \url{https://arxiv.org/abs/2302.03337}
\BIBentrySTDinterwordspacing

\bibitem{aft-ai-traffic-00}
\BIBentryALTinterwordspacing
A.~Fressancourt, L.~Iannone, Z.~Lou, and D.~Trossen, ``{Handling inter-DC/Edge
  AI-related network traffic: Problem statement},'' Internet Engineering Task
  Force, Internet-Draft draft-aft-ai-traffic-00, oct 2024. [Online]. Available:
  \url{https://datatracker.ietf.org/doc/draft-aft-ai-traffic/00/}
\BIBentrySTDinterwordspacing

\bibitem{ouyang2021communication}
S.~Ouyang, D.~Dong, Y.~Xu, and L.~Xiao, ``Communication optimization strategies
  for distributed deep neural network training: {A} survey,'' \emph{Journal of
  Parallel and Distributed Computing}, vol. 149, 2021.

\bibitem{cao2024network}
P.~Cao, W.~Cheng, S.~Zhao, and Y.~Xiong, ``Network load balancing with parallel
  flowlets for ai training clusters,'' in \emph{Proc. in {SIGCOMM} Workshop on
  Networks for AI Computing}, Aug 2024.

\bibitem{thaler2000multipath}
D.~Thaler and C.~Hopps, ``Multipath issues in unicast and multicast next-hop
  selection,'' Tech. Rep., 2000.

\bibitem{vanini2017let}
E.~Vanini, R.~Pan, M.~Alizadeh, P.~Taheri, and T.~Edsall, ``Let it flow:
  {Resilient} asymmetric load balancing with flowlet switching,'' in
  \emph{Proc. USENIX Symposium on Networked Systems Design and Implementation
  ({NSDI})}, March 2017.

\bibitem{huang2021mitigating}
J.~Huang, W.~Lyu, W.~Li, J.~Wang, and T.~He, ``Mitigating packet reordering for
  random packet spraying in data center networks,'' \emph{IEEE/ACM Transactions
  on Networking}, vol.~29, no.~3, 2021.

\bibitem{259355}
M.~Al-Fares, S.~Radhakrishnan, B.~Raghavan, N.~Huang, and A.~Vahdat,
  ``{Hedera}: {Dynamic} flow scheduling for data center networks,'' in
  \emph{proc. {USENIX} Symposium on Networked Systems Design and Implementation
  ({NSDI})}, Apr 2010.

\bibitem{ultraethernet}
U.~E. Consortium, ``The new era needs a new network,''
  \url{https://ultraethernet.org/}, 2008, [Online; accessed 18-Feb-2025].

\bibitem{metz2024empowering}
J.~Metz, ``Empowering ai workloads in ultra ethernet consortium,'' in
  \emph{Proc. {IEEE} Photonics Society Summer Topicals Meeting Series ({SUM})},
  July 2024.

\bibitem{dixit2013impact}
A.~Dixit, P.~Prakash, Y.~C. Hu, and R.~R. Kompella, ``On the impact of packet
  spraying in data center networks,'' in \emph{Proc. {IEEE} International
  Conference on Computer Communications {(INFOCOM}}, Apr 2013.

\bibitem{ghorbani2017drill}
S.~Ghorbani, Z.~Yang, P.~B. Godfrey, Y.~Ganjali, and A.~Firoozshahian,
  ``{Drill}: {Micro} load balancing for low-latency data center networks,'' in
  \emph{Proc. {ACM} Special Interest Group on Data Communication ({SIGCOMM})},
  Aug 2017.

\bibitem{le2024strack}
Y.~Le, R.~Pan, P.~Newman, J.~Blendin, A.~Kabbani, V.~Jain, R.~Sivaramu, and
  F.~Matus, ``{STrack}:{A} reliable multipath transport for ai/ml clusters,''
  \emph{arXiv preprint arXiv:2407.15266}, 2024.

\bibitem{floyd1993random}
S.~Floyd and V.~Jacobson, ``Random early detection gateways for congestion
  avoidance,'' \emph{IEEE/ACM Transactions on networking}, vol.~1, no.~4, 1993.

\bibitem{handley2017re}
M.~Handley, C.~Raiciu, A.~Agache, A.~Voinescu, A.~W. Moore, G.~Antichi, and
  M.~W{\'o}jcik, ``Re-architecting datacenter networks and stacks for low
  latency and high performance,'' in \emph{Proc. {ACM} Special Interest Group
  on Data Communication ({SIGCOMM})}, Aug 2017.

\bibitem{lu2018multi}
Y.~Lu, G.~Chen, B.~Li, K.~Tan, Y.~Xiong, P.~Cheng, J.~Zhang, E.~Chen, and
  T.~Moscibroda, ``{Multi-Path} transport for {RDMA} in datacenters,'' in
  \emph{Proc. USENIX symposium on networked systems design and implementation
  ({NSDI})}, Apr 2018.

\bibitem{qureshi2022plb}
M.~A. Qureshi, Y.~Cheng, Q.~Yin, Q.~Fu, G.~Kumar, M.~Moshref, J.~Yan,
  V.~Jacobson, D.~Wetherall, and A.~Kabbani, ``{PLB}: {Congestion} signals are
  simple and effective for network load balancing,'' in \emph{Proc. {ACM}
  Special Interest Group on Data Communication ({SIGCOMM})}, Aug 2022.

\bibitem{9269053}
J.~Lim, S.~Nam, J.-H. Yoo, and J.~W.-K. Hong, ``Best nexthop load balancing
  algorithm with inband network telemetry,'' in \emph{Proc. {IEEE}
  International Conference on Network and Service Management (CNSM)}, Nov 2020.

\bibitem{reps2025}
T.~Bonato, A.~Kabbani, A.~Ghalayini, M.~Papamichael, M.~Dohadwala,
  L.~Gianinazzi, M.~Khalilov, E.~Achermann, D.~De~Sensi, and T.~Hoefler,
  ``{REPS}: {Recycled} entropy packet spraying for adaptive load balancing and
  failure mitigation,'' \emph{arXiv preprint arXiv:2407.21625v3}, 2025.

\bibitem{cao2013per}
J.~Cao, R.~Xia, P.~Yang, C.~Guo, G.~Lu, L.~Yuan, Y.~Zheng, H.~Wu, Y.~Xiong, and
  D.~Maltz, ``Per-packet load-balanced, low-latency routing for clos-based data
  center networks,'' in \emph{Proc. {ACM} conference on Emerging networking
  experiments and technologies ({CoNEXT})}, Dec 2013.

\bibitem{ramakrishnan2001addition}
K.~Ramakrishnan, S.~Floyd, and D.~Black, ``The addition of explicit congestion
  notification ({ECN}) to {IP},'' Tech. Rep., 2001.

\bibitem{wu2012tuning}
H.~Wu, J.~Ju, G.~Lu, C.~Guo, Y.~Xiong, and Y.~Zhang, ``Tuning {ECN} for data
  center networks,'' in \emph{Proc. of the international conference on Emerging
  networking experiments and technologies (CoNECT)}, Dec 2012.

\bibitem{fisheryates}
M.~Eberl, ``Fisher–yates shuffle,'' \url{https://ultraethernet.org/}, 2016,
  [Online; accessed 28-Feb-2025].

\bibitem{luo2024seqbalance}
H.~Luo, J.~Zhang, M.~Yu, Y.~Pan, T.~Pan, and T.~Huang, ``{SeqBalance}:
  {Congestion}-aware load balancing with no reordering for roce,'' \emph{arXiv
  preprint arXiv:2407.09808}, 2024.

\bibitem{BFD}
S.~Hongye, ``What is bfd?''
  \url{https://info.support.huawei.com/info-finder/encyclopedia/en/BFD.html},
  2024, [Online; accessed 31-Mar-2025].

\bibitem{dpfr}
Y.~Xiaoli, ``What is dpfr?''
  \url{https://info.support.huawei.com/info-finder/encyclopedia/en/DPFR.html},
  2024, [Online; accessed 31-Mar-2025].

\bibitem{broadcom}
P.~D.~V. Mohan~Kalkunte, Niranjan~Vaidya, ``Cognitive routing in the tomahawk 5
  data center switch,'' \url{
  https://www.broadcom.com/blog/cognitive-routing-in-the-tomahawk-5-data-center-switch},
  2023, [Online; accessed 31-Mar-2025].

\bibitem{benson2011microte}
T.~Benson, A.~Anand, A.~Akella, and M.~Zhang, ``{MicroTE}: Fine grained traffic
  engineering for data centers,'' in \emph{Proc. of the seventh conference on
  emerging networking experiments and technologies(CoNEXT)}, Dec 2011.

\bibitem{curtis2011mahout}
A.~R. Curtis, W.~Kim, and P.~Yalagandula, ``{Mahout}: Low-overhead datacenter
  traffic management using end-host-based elephant detection,'' in \emph{Proc.
  IEEE International Conference on Computer Communications (INFOCOM)}, Apr
  2011.

\bibitem{rasley2014planck}
J.~Rasley, B.~Stephens, C.~Dixon, E.~Rozner, W.~Felter, K.~Agarwal, J.~Carter,
  and R.~Fonseca, ``{Planck}: {Millisecond}-scale monitoring and control for
  commodity networks,'' \emph{ACM SIGCOMM Computer Communication Review},
  vol.~44, no.~4, 2014.

\bibitem{kabbani2014flowbender}
A.~Kabbani, B.~Vamanan, J.~Hasan, and F.~Duchene, ``{Flowbender}: {Flow}-level
  adaptive routing for improved latency and throughput in datacenter
  networks,'' in \emph{Proc. {ACM} International on Conference on emerging
  Networking Experiments and Technologies (CoNEXT)}, Dec 2014.

\bibitem{katta2016hula}
N.~Katta, M.~Hira, C.~Kim, A.~Sivaraman, and J.~Rexford, ``{Hula}: {Scalable}
  load balancing using programmable data planes,'' in \emph{Proc. Symposium on
  {SDN} Research ({SOSR})}, Mar 2016.

\bibitem{katta2016clove}
N.~Katta, M.~Hira, A.~Ghag, C.~Kim, I.~Keslassy, and J.~Rexford, ``{CLOVE}:
  {How} i learned to stop worrying about the core and love the edge,'' in
  \emph{Proc. {ACM} Workshop on Hot Topics in Networks($HotNets$)}, Nov 2016.

\bibitem{he2015presto}
K.~He, E.~Rozner, K.~Agarwal, W.~Felter, J.~Carter, and A.~Akella, ``{Presto}:
  Edge-based load balancing for fast datacenter networks,'' \emph{ACM SIGCOMM
  Computer Communication Review}, vol.~45, no.~4, 2015.

\bibitem{perry2014fastpass}
J.~Perry, A.~Ousterhout, H.~Balakrishnan, D.~Shah, and H.~Fugal, ``{Fastpass}:
  {A} centralized" zero-queue" datacenter network,'' in \emph{Proc. ACM Special
  Interest Group on Data Communication {SIGCOMM}}, Aug 2014.

\bibitem{huang2018qdaps}
J.~Huang, W.~Lv, W.~Li, J.~Wang, and T.~He, ``{QDAPS}: {Queueing} delay aware
  packet spraying for load balancing in data center,'' in \emph{Proc.
  International Conference on Network Protocols ({ICNP})}.\hskip 1em plus 0.5em
  minus 0.4em\relax IEEE, Sep 2018.

\bibitem{liu2019ag}
J.~Liu, J.~Huang, W.~Li, and J.~Wang, ``{AG}: {Adaptive} switching granularity
  for load balancing with asymmetric topology in data center network,'' in
  \emph{Proc. International Conference on Network Protocols ({ICNP})}, Oct
  2019.

\bibitem{csig}
N.~M. J.~K. A.~Ravi, N.~Dukkipati, ``Congestion signaling ({CSIG}),''
  \url{https://www.ietf.org/archive/id/draft-ravi-ippm-csig-01.html}, 2024,
  [Online; accessed 31-Mar-2025].

\bibitem{int}
T.~P. A.~W. Group, ``In-band network telemetry ({INT}) dataplane
  specification,'' \url{https://p4.org/p4-spec/docs/INT_v2_1.pdf}, 2020,
  [Online; accessed 31-Mar-2025].

\end{thebibliography}

\end{document}